\documentclass[aps,prd,reprint,onecolumn,notitlepage,superscriptaddress]{revtex4-1}

\usepackage{tcolorbox}
\usepackage{xcolor}
\usepackage{soul}
\usepackage{lipsum}
\usepackage{tikz}
\usepackage[compat=1.0.0]{tikz-feynman}
\usetikzlibrary{shapes}
\usetikzlibrary{positioning}
\usepackage{graphicx}
\usepackage{subfig}
\usepackage{color}
\usepackage{appendix}
\usepackage{mathtools, nccmath, amssymb}
\usepackage{verbatim}
\usepackage{hyperref}
\usepackage{bm}
\usepackage[mathscr] {euscript}
\newcommand{\rom}[1]{\textup{\uppercase\expandafter{\romannumeral#1}}}

\allowdisplaybreaks

\begin{document}

\title{Revisiting Lagrangian Formulation of Stochastic inflation}

\author{Rajat Kumar Panda \footnote{email:rajat23@iiserb.ac.in}
       }
       
\affiliation{Indian Institute of Science Education and Research Bhopal,\\ Bhopal 462066, India}       
\author{Sukanta Panda \footnote{email:sukanta@iiserb.ac.in}
       }

\affiliation{Indian Institute of Science Education and Research Bhopal,\\ Bhopal 462066, India}

\author{Abbas Tinwala \footnote{email:abbas@prl.res.in}
       }

\affiliation{Physical Research Laboratory,\\ Ahmedabad 380009, India}

\begin{abstract}
We revisit the Lagrangian formulation of stochastic inflation, where the path-integral approach is employed to derive the Langevin equation governing the dynamics of long-wavelength fields, in contrast to the standard method where the Langevin equation is derived directly from the equation of motion of the full quantum field. Focusing on a massless, minimally coupled scalar field with quartic self-interaction in a de Sitter background, we re-derive the formal expression for the influence functional that encapsulates the effects of short-wavelength fields up to second order in the coupling constant, and compare our results with those obtained in earlier works. In doing so, we highlight certain subtleties that have been previously overlooked, including the non-orthogonality between long- and short-wavelength modes, which we analyze in detail, as well as the absence of a consistent prescription for handling general interaction terms in the imaginary part of the influence functional. The latter issue points to a broader challenge: the lack of a universally accepted framework for treating the imaginary component of effective actions.
\end{abstract}



\maketitle

\section{Introduction}\label{Introduction}
    The theory of inflation has become a cornerstone of modern cosmology since it's introduction by Guth in 1981 \cite{Guth:1980zm}, providing elegant resolutions to explain several profound puzzles inherent in the standard hot big bang framework like the horizon and flatness problems. By proposing a brief epoch of accelerated expansion in the early universe, inflation accounts for the observed large scale homogeneity, isotropy and spatial flatness naturally. At the same time, it also provides a mechanism for generating the primordial fluctuations that seeded the large-scale structures we observe today \cite{Linde:1981mu,Albrecht:1982wi,Mukhanov:1981xt,Hawking:1982my,Starobinsky:1982ee,Guth:1982ec}. In its simplest formulation, inflation is driven by a scalar degree of freedom, the inflaton rolling slowly along a nearly flat potential. Quantum fluctuations of this scalar field, stretched to super Hubble scales by accelerated expansion ultimately give rise to classical density perturbations in the post inflationary era \cite{Linde:1983gd}.\\
Although this treatment of inflation, driven by a homogeneous scalar field, does capture the essential dynamics of accelerated expansion, there are compelling reasons to move beyond this picture. Inflation is an intrinsically time dependent, non-equilibrium process and it demands descriptions that properly account for phenomena like decoherence \cite{Morikawa:1989xz,Kiefer:1998jk}, the effective dynamics of super-Hubble modes in presence of continuous horizon crossing by sub-horizon modes to become super-horizon ones among others. For example, the quantum-to-classical transition of scalar field perturbations cannot be naturally addressed within a purely deterministic framework . These issues motivated the development of stochastic inflation formalism, where the long-wavelength modes of the inflation are treated as coarse-grained, effectively classical fields subjected to stochastic noise sourced by the continuous horizon-crossing of quantum fluctuations\cite{Starobinsky:1986fx,Nambu:1988je} . The effective equations of motion come out as a Langevin type equation having an inherent random noise dependence and through this, stochastic inflation establishes a bridge between microscopic quantum fluctuations and macroscopic cosmological observables which are classical in nature\cite{Burgess:2010dd,Vennin:2015hra}.\\
But to place stochastic inflation on firm theoretical footing, it is essential to derive it from first principles, which is often done by extremizing the action in physics and hence getting equations of motion, which should be noise dependent here because of the exact procedures we did during extremizing the action rather than introducing noise by hand in the equations of motion. One way to do this is known as the influence functional method, originally developed by Feynman and Vernon to describe open quantum systems\cite{Feynman:1963fq}. Within this approach,  by treating the long-wavelength modes of the inflaton as the system of interest, the short-wavelength modes can be systematically integrated out leaving an influence functional term in the effective action for the long-wavelength modes that encode both dissipative and stochastic terms\cite{Calzetta:1993qe,Calzetta:2008iqa}. Through this influence functional, short-wavelength modes affect the effective action for long-wavelength modes and hence the dynamics of the long-wavelength modes. Of particular interest is the imaginary part of this influence functional and the subject of how to interpret it. It turns out that through a particular transformation known as the Hubbard-Stratonovich (HS) transformation \cite{1957SPhD2416S, Hubbard:1959ub}, one can recast the imaginary part of this influence functional back into a path integral over a classical random field weighted by a real probability distribution (unlike the usual oscillatory weight $\sim e^{iS}$ in the path integral formulation where $S$ is the real action in QFT). This procedure leads to a Langevin-type equation of motion for the long-wavelength fields first obtained in \cite{Morikawa:1989xz} (actually Starobinsky derived the Langevin equation for the first time in \cite{Starobinsky:1986fx} although he did not use the lagrangian formulation) where the random noise terms follow the statistics determined by the probability distribution function obtained through the HS transformation \footnote{It is important to note however, that Starobinsky obtained the Langevin equation directly by splitting the full field into long and short-wavelength modes in the equation of motion for the full field. See also \cite{Vennin:2020kng} and other references therein where this way of obtaining the Langevin equation is followed}. Hence, unlike in standard treatment where the superhorizon modes freeze out once they cross the horizon, in case of stochastic inflation, their dynamics does get affected, albeit by a random noise term, which in itself depends upon the small-wavelength modes. The Langevin equation can be solved and the correlation functions of the long-wavelength fields can be computed using the correlation functions of various noise variables that appear in the equation of motion. This procedure is known to capture all the leading infrared logarithms that appear in the correlation function of the quantum field calculated using quantum field theoretical formulation in de Sitter space (see \cite{Tsamis:2005hd,Onemli:2015pma} and other references therein).\\
Let us briefly describe our article. In Sec.\ref{Sec2} we will derive a formal expression for the influence functional for scalar field theory in exact de Sitter space with a quartic-self interaction. This is achieved by splitting the full quantum field into short and long-wavelength parts in the action and performing a path integral over the short-wavelength fields. Since we are interested in calculating the in-in expectation values of the long-wavelength fields for out-of-equilibrium quantum field theory as opposed to the
usual in-out transition amplitude, we will use the Schwinger-Keldysh Closed Time Path (CPT) formalism \cite{Schwinger:1960qe, Keldysh:1964ud,Calzetta:1986ey} (on which we have provided brief details in Sec.\ref{Sec2}).\\ 
Although such a derivation was carried out in the past in \cite{PerreaultLevasseur:2013kfq}, we do it here again to fill in some gaps left out previously. The calculation performed in this paper yields additional terms in the influence functional when compared with \cite{PerreaultLevasseur:2013kfq}. There are two reasons why the new terms appear in our calculation. First one is associated with the subject of orthogonality of the long and short-wavelength fields. As we will show in Sec.\ref{Sec4}, a spatial integral of only a term bilinear in long and short-wavelength field can be zero when a step function is chosen as the window function to split the full field. A spatial integral of any other term which is not linear in either of the short or long-wavelength fields is not strictly zero and becomes the source for the new terms in the influence functional. We believe this issue is subtle and the ignorance of the additional terms arising in this way has not been well reasoned in the past. Although we have good reason to expect these terms from a mathematical viewpoint, we lack any explanation from a physical perspective as to why these terms should arise. It may very well be the case that these terms are just unwanted artifacts of the procedure followed to perform the calculations. It is to be noted that the new terms arising in this way are a feature of an interacting theory only and therefore the results from free theory derived in Sec.\ref{Sec3} do match with those obtained earlier in \cite{Morikawa:1989xz, Matarrese:2003ye}. The second reason has to do with the interaction term linear in the short-wavelength field that arises after we split the full field into short and long-wavelength fields in the action. Such a term acts like a source in the action producing additional diagrams in the influence functional. Usually when path integral formalism is used one adds an auxiliary source term in the classical action which is linear in the quantum field to deal with the interaction terms by which we mean all terms which are higher than quadratic in power of the quantum field. Such terms in $e^{iS}$ are perturbatively expanded in powers of coupling constants and expressed as functional derivatives of $e^{iS_{\text{free}}}$ with respect to the source. At the end of the calculation the source is then taken to zero since it was just a trick to deal with higher-order terms. However, in our case the source is real, not auxiliary and therefore additional terms that depend on this source cannot be taken zero in the end. As per our analysis, these terms bear no mention of their existence or the reason for their removal in \cite{PerreaultLevasseur:2013kfq}. As we will show in Sec.\ref{Sec5} these additional terms bring with them the difficulty of interpreting their imaginary contribution to the influence functional because they bear a form which renders the HS transformation inapplicable. This issue points towards a more general problem of how to deal with the imaginary part of the influence functional. As per our knowledge this problem remains unresolved making the subject quite interesting to study at least from a theoretical viewpoint. Sec.\ref{Conclusion} bears a summary of our study.\\
Throughout this paper we will be using the following conventions and notations which resembles that due to DeWitt's. We reserve the greek letter `$\varphi^i$' for the long-wavelength fields where the discrete field index and the field's spacetime argument are condensed into the single label $i$. For instance, $\varphi^i$ is equivalent to $\varphi^I(x)$, where capital Latin letters $(I, J,...)$ are used as a placeholder for conventional field indices. Thus, the small Latin indices such as `$i$' stands for the pair $(I,x)$. Additionally, the following summation convention is used in four dimensions.

\begin{equation}\label{2}
    \varphi^i B_{ij} \varphi^j = \int dt_x \ a^3(t_x)\int dt_{x'} \ a^3(t_{x'}) \int d^3x \int d^3x' \ \varphi^I(x) B_{IJ}(x,x') \varphi^J(x'),
\end{equation}

\section{Effective Action for the long-wavelength field}\label{Sec2}

We begin with the action for a scalar field $\Phi$ with quartic interaction term in the de Sitter spacetime. 

\begin{align}\label{action}
    S = \int d^4x \sqrt{-g(x)}\left(\frac{1}{2}g_{\mu\nu}\nabla_\mu\Phi\nabla_\nu\Phi - \frac{1}{4}\lambda\Phi^4\right).
\end{align}

The background metric is given by

\begin{align}\label{metric}
    ds^2 = -dt^2 + a^3(t)\overrightarrow{dx}\cdot\overrightarrow{dx},
\end{align}

where $a(t)$ is the scale factor of the expanding universe. Throughout this paper, we will assume the Hubble paramater defined by $H = \frac{\dot a}{a}$ to be fixed. Additionally, we won't be considering metric perturbations in this work.\\
We split the quantum field $\Phi$ into two pieces. One containing the long-wavelength fields and the other short ones as follows

\begin{align}\label{split}
    \Phi = \varphi + \psi.
\end{align}

The split is made using a window function in momentum space as follows,

\begin{align}\label{short} 
    \psi(x) = \int\frac{d^3k}{(2\pi)^3}W(k,t)(\phi_\mathbf{k}\hat a_\mathbf{k} e^{-i\mathbf{k}\cdot\mathbf{x}} + \phi^*_\mathbf{k}\hat a^\dagger_\mathbf{k} e^{i\mathbf{k}\cdot\mathbf{x}}).
\end{align}

where $\phi_\mathbf{k}$ are the mode solutions of the Euler-Lagrange equations of motion satisfied by $\Phi$ with the annihilation operator $\hat a_\mathbf{k}$ annihilating the Bunch-Davies vacuum state. The window function $W(k,t)$ acts like a high-pass filter that would allow only those wavelengths higher than a certain cut-off. The long-wavelength fields denoted by $\varphi$ are also called ``coarse-grained" fields since they do not contain the fine details of the full field at comoving wavenumbers larger than the chosen cut-off. For this reason the cutoff is more appropriately called the coarse-graining scale or the smoothing scale. Throughout this paper we will be using the term ``long-wavelength fields" for $\varphi$. 

There are two questions regarding the window function. First is what should be the cut-off to separate the long-wavelength (infrared fields) from the short-wavelength fields (UV fields). It makes sense to choose a de Sitter invariant physical cut-off, just like one must choose an invariant regularization scheme for isolating divergences in general curved spacetime \cite{Tsamis:2005hd}. Moreover the interesting part of the study of a scalar field evolving in de Sitter space is the particle production \cite{Parker:1969au} that becomes significant when $k<a(t)H$ \cite{Tsamis:2005hd}. For these reasons, a suitable choice for the physical cut-off is the de Sitter invariant length $H^{-1}$. The second question is what form of window function should we choose which has been a subject of discussion in the past work \cite{Winitzki:1999ve}. The simplest of all is the sharp cut-off Heaviside theta function $\theta(k-\sigma aH)$ chosen by Starobinsky \cite{Starobinsky:1986fx}, where $\sigma$ is called the coarse-graining parameter, taken to be a small real number much less than 1 (we will discuss the significance of this paramater in a moment). This choice is special in the sense that the calculations become easy and for massless minimally coupled free scalars evolving in pure de Sitter space it leads to white noise. This makes the evolution of the coarse-grained fields a Markovian process, in which the instantaneous evolution of the coarse-grained field is governed by the statistics followed by the noise only at that instant of time. The calculation is simplified in this case because one does not require to know what values the noise variables could have assumed prior to the ``time" at which the equation of the coarse-grained fields is solved. However, it was shown in \cite{Winitzki:1999ve} that this choice does not correctly reproduce the behavior of two-point correlation of the time-differentiated long-wavelength fields $\langle\dot\varphi(\mathbf{x},t_x)\dot\varphi(\mathbf{y},t_y)\rangle$ at large spatial separation $r=|\mathbf{x}-\mathbf{y}|$. It was further shown there, that any smooth cut-off function that satisfies certain basic properties does not lead to above-mentioned unusual behavior of the correlation function. A coarse-graining procedure that uses a general class of exponential filters satisfying the properties mentioned in \cite{Winitzki:1999ve} can be found in \cite{Mahbub:2022osb}. However, we will only be calculating correlation functions with coincident spatial arguments in this paper and so we will stick with the Heaviside theta function as our choice for the window function. \\
The most interesting feature of the window function is its time dependence which is a direct consequence of using comoving coordinates in exact de Sitter space (note that the physical cutoff would be $k_\text{phys} = \sigma H$ which is time-independent for exact de Sitter). Indeed, the whole technique of treating the collective effect of the short-wavelength fields as stochastic force in the Langevin-type equation for the long-wavelength fields hinges on the time dependence of the window function. In simple words, the time dependence of the window function leads to the crossing of the modes from the sub-Hubble regime to the super-Hubble regime with time. Since the fields under consideration are fundamentally quantum in nature they affect the dynamics of the long-wavelength fields in a random manner. The question is how does this affect the nature of the long-wavelength fields. It is true that both long-wavelength and short-wavelength fields are quantum in nature by definition (\ref{split}). However, as explained in detail in \cite{Polarski:1995jg, Lesgourgues:1996jc, Kiefer:1998qe, Albrecht:1992kf}, after horizon exit, the decaying mode of the full quantum field can be neglected leading to loss of coherence between the decaying and the non-decaying modes. The consequence of this is that the long-wavelength fields (the full quantum field which has crossed the horizon) looses its quantum nature that was preserved in its commutation relation with its canonical conjugate field and becomes a classical field albeit stochastic in nature. The classicalization of the quantum field after horizon crossing in this manner was called ``decoherence without decoherence" in \cite{Polarski:1995jg} because it does not rely on interactions of the scalar field with other fields. The immediate usefulness of this process is that the quantum expectation values of the long-wavelength fields turn into statistical averages which can be calculated using the probability distribution of the noise terms. We will make use of this technique to compute the two-point correlation of the long-wavelength fields in the next sections.\\ 
As mentioned in the introduction, we are interested in calculating the in-in expectation values for which the standard method to use is the Closed-Time-Path (CTP) formalism due to Schwinger and Keldysh \cite{Schwinger:1960qe, Keldysh:1964ud, Calzetta:1986ey, Calzetta:2008iqa}. It does not make sense to use the usual path integral formalism of ordinary quantum field theory for this purpose because we are interested in studying how the expectation values evolves with time and so we do not presume to know what the final state or out-state for the system is. This is precisely what the CTP formalism achieves because it differs in the way it is constructed as compared to the usual path integral formalism (see \cite{Jordan:1986ug}). In CTP formalism, one introduces two paths, one running forward in time from $t=t_i$ to $t=t_f$ called $\mathcal{P}_+$ and the other running backward in time from $t=t_f$ to $t=t_i$ called $\mathcal{P}_-$ along with two different sources $J_+$ and $J_-$ associated with the two paths which are a priori assumed to be different and independent. The fields are then assumed to evolve forward in time on $\mathcal{P}_+$ in the presence of $J_+$ and then backward in time in presence of the source $J_-$ with the condition that they match at $t=t_f$. The closed-time-path contour so formed can then be converted back to usual single-time path in the following way. Denoting by $\Phi_+$ the fields evolving on the forward path and by $\Phi_-$ for the fields evolving on the backward path the generating functional, $\mathscr{Z} = e^{iW}$ for the CTP formalism can be written as 

\begin{align}
    \mathscr{Z} &= \int \mathscr{D}\Phi_+\int\mathscr{D}\Phi_-\exp\{i(S_+[\Phi_+]-S_-[\Phi_-] + \int d^4x \ a^3(t_x)(\  J_+\Phi_+-J_-\Phi_-)\}\Big|_{\Phi_+(t_f)=\Phi_-(t_f)}\nonumber\\
    &=\int_{-\infty}^\infty d\Phi_+(t_N)\int_{-\infty}^\infty d\Phi_-(t_N) \ \delta(\Phi_+(t_N)-\Phi_-(t_N))\int_{-\infty}^\infty d\Phi_+(t_{N-1})\int_{-\infty}^\infty d\Phi_-(t_{N-1}) \ ... \nonumber\\
    &\times\int_{-\infty}^\infty d\Phi_+(t_0)\int_{-\infty}^\infty d\Phi_-(t_0)\prod_{i=0}^N\exp\left[i\Delta t\Big\{\mathcal{L}_+[\Phi_+(t_i)]-\mathcal{L}_-[\Phi_-(t_i)] + a^3(t_i)\int d^3\mathbf{x} \Big(J_+(t_i)\Phi_+(t_i) - J_-(t_i)\Phi_-(t_i)\Big)\Big\}\right],
\end{align}

where we have discretized the time coordinate such that $t_N=t_f$, $t_0=t_i$ and $N\Delta t = t_f-t_i$. The boundary condition has been taken into consideration using the $\delta$ function in the second step. If functional derivatives of $W[J_+,J_-]$ are now computed wrt. either $J_+$ or $J_-$ and $J_+=J_-$ used at the end of the calculation one gets the desired in-in expectation values of the field operators \cite{Jordan:1986ug}.

Since we are interested in deriving the effective action of the long-wavelength fields we will only perform the closed-time-path integral over the short-wavelength. For this, we make the split (\ref{split}) using (\ref{short}) in the action for the full field given in (\ref{action}) to obtain

\begin{align}
    S[\Phi]= S[\varphi]+S[\psi]+S_{\text{int}} [\varphi,\psi],
\end{align}
    
where all the terms depending both on $\varphi$ and $\psi$ have been collected in $S_\text{int}$. The generating functional in the CTP formalism reads

\begin{align}
    \mathscr {Z} = \int \mathscr{D} \varphi^+ \mathscr{D} \varphi^- \int_{\text{B.C.}} \mathscr{D} \psi^+ \mathscr{D} \psi^- \text{exp}\left\{i\left[S_+\left(\Phi_+\right)-S_-\left(\Phi_-\right)\right]\right\},
\end{align}

where B.C. stands for the boundary condition satisfied by the short-wavelength fields. We are interested in the reduced generating functional denoted by $Z$ and defined by

\begin{align}
        Z[\varphi\pm] &= \int_\text{B.C.} \mathscr{D} \psi^+ \mathscr{D} \psi^- \exp{i[S^+(\psi^+)-S^-(\psi^-)+S_{\text{int}} [\varphi^+,\psi^+] - S_{\text{int}} [\varphi^-,\psi^-]]}\nonumber\\
        &=e^{iS_{\text{inf}}[\varphi_\pm]},
\end{align}

where we have also defined the influence functional denoted by $S_{\text{inf}}$ that contains the complete ``influence" due to the short-wavelength fields. The effective action $\Gamma$ of the long-wavelength fields is then defined by

\begin{align}
    \Gamma[\varphi_\pm] = S_+[\varphi_+] - S_-[\varphi_-] + S_{\text{inf}}[\varphi_\pm]. 
\end{align}

The generating functional written down in terms of the effective action then reads

\begin{align}
    \mathscr{Z} &= \int\mathscr{D}\varphi_+\int\mathscr{D}\varphi_-\exp\{i(S_+[\varphi_+]-S_-[\varphi_-]\}\times Z[\varphi_\pm]\\
    &=\int\mathscr{D}\varphi_+\int\mathscr{D}\varphi_-\exp\{i(S_+[\varphi_+]-S_-[\varphi_-] + S_\text{inf}[\varphi_\pm]\}\\
    &=\int\mathscr{D}\varphi_+\int\mathscr{D}\varphi_-e^{i\Gamma[\varphi_\pm]}.
\end{align}

Let us now perform computations for the theory given in (\ref{action}). Using the expression for the metric in (\ref{metric}) the action can be written as 

\begin{align}
        S=\int d^4x \left\{ \frac{1}{2}\Phi(x)\Lambda(x)\Phi(x)-V(\Phi)\right\},
\end{align}
    
where $\Lambda$ is an operator which is second-order in space-time derivatives defined by
        
\begin{align}
    \Lambda=-a^3(t)\left(\frac{\partial^2}{{\partial t}^2}+3H\frac{\partial}{\partial t}- \frac{\nabla^2}{a^2}+m^2\right).
\end{align}
    
The reduced generating functional obtained after expanding the full field $\Phi$ around the long-wavelength fields in the action is
    \begin{align}
        Z[\varphi_\pm]=\left(\prod_{i=+,-}\int \mathscr{D}\psi_i\right) \ \exp\left[i\left\{\frac{1}{2}\psi_i\eta_i\Lambda_i\psi_i + \psi_i\eta_i\Lambda_i\varphi_i-\sum_{N=1}^\infty \frac{1}{N!}V_N[\varphi_i]\psi_i^N\eta_i\right\}\right],
    \end{align}
    where $V_N = \frac{\delta^N V[\varphi]}{\delta\varphi^N}$. Note that we have used the DeWitt's notation here. The small latin letters $i,j...$ are the condensed indices which stand for $(\pm,x)$ and the capital latin letters $I,J,..$ (which will appear later) would stand for the contour labels $\pm$.
    
    For a scalar field with quartic interaction we have $V[\Phi]=\frac{\lambda}{4!}\Phi^4$ and thus we obtain the following expression for the reduced generating functional:
        \begin{align}
            Z[\varphi_\pm]=\int \mathscr{D}\psi \exp{\left[i\left\{\frac{1}{2}\psi_i\Lambda_i\psi_i\eta_i+\psi_i\Lambda_i\varphi_i\eta_i-\frac{\lambda\varphi_i^3}{3!}\psi_i\eta_i-\frac{1}{2}\left(\frac{\lambda\varphi_i^2}{2}\right)\psi_i^2\eta_i-\frac{1}{3!}\lambda\varphi_i\psi_i^3\eta_i-\frac{\lambda}{4!}\psi_i^4\eta_i\right\}\right]}.
        \end{align}
        It is convenient to use the redefined operator 
        \begin{align}
         E_i=\Lambda_i-\left(\frac{\lambda\varphi_i^3}{3!}\right),
        \end{align}
    so that
        \begin{align}
            Z[\varphi_\pm]= \int \mathscr{D}\psi \exp{\left[i\left\{\frac{1}{2}\psi_i \Lambda_i\psi_i\eta_i+\psi_i E_i\varphi_i\eta_i-\frac{1}{4}\lambda\varphi_i^2\psi_i^2\eta_i-\frac{1}{3!}\lambda\varphi_i\psi_i^3\eta_i-\frac{\lambda}{4!}\psi_i^4\eta_i\right\}\right]}.
        \end{align}
     The term linear in $\psi$ in the expression above is like the source term. It is not auxiliary and hence cannot be put to zero as is usually done in the path integral formalism. If we define $J_i=E_i\varphi_i$, we can express the complete reduced generating functional in terms of the ``free" generating functional using functional derivatives wrt. to the source $J_i$ as follows:
        \begin{align} \label{genfunc}
            Z[\varphi_\pm]= \exp{\left\{-\frac{i\lambda}{4}\varphi_i^2\eta_i\left(\frac{1}{i\eta_i}\frac{\delta}{\delta J_i}\right)^2\right\}}\exp{\left\{-\frac{i\lambda}{3!}\varphi_i\eta_i\left(\frac{1}{i\eta_i}\frac{\delta}{\delta J_i}\right)^3\right\}}\exp{\left\{-\frac{i\lambda}{4!}\eta_i\left(\frac{1}{i\eta_i}\frac{\delta}{\delta J_i}\right)^4\right\}} Z_\text{f}[\varphi_\pm;J_\pm],
        \end{align}
    where the ``free" reduced generating functional is defined by
        \begin{align}
            Z_\text{f}[\varphi\pm;J_\pm]= \left(\prod_{i}\int \mathscr{D}\psi_i\right) \exp{\left[i\left\{\frac{1}{2}\psi_i D_i\psi_i\eta_i+\psi_i J_i\eta_i\right\}\right]}.
        \end{align}
    After performing the Gaussian integral in the expression above we obtain
    
    \begin{align} \label{freegenfunc}
        Z_\text{f}[\varphi_\pm;J_\pm]= N \exp{\left\{-\frac{i}{2}\eta_i J_i G_{ij} J_j\eta_j\right\}},    \end{align}
    
    where the Green's function $G_{ij}$ satisfies $\Lambda_iG_{ij}=\delta_{ij}$. Here, $\delta_{ij} = \delta(x-y)\delta_{IJ}$. The matrix $G_{ij}$ has the following expression:
    
\begin{align}\label{greensfunc}
    G_{ij} = G_{IJ}(x,y) \equiv-i
    \begin{pmatrix}
        \langle T[\psi_+(x)\psi_+(y)]\rangle & \langle \psi_-(y)\psi_+(x)\rangle \\\langle \psi_-(x)\psi_+(y)\rangle &\langle \bar T[\psi_-(x)\psi_-(y)]\rangle
    \end{pmatrix}.
\end{align}

The upper-left element of $G_{ij}$ is the usual time-ordered correlation of fields lying on the forward contour. The lower-right element is the anti-time-ordered correlation of fields lying on the backward branch of the counter. The anti-time ordering appears because, on the backward branch of the contour, the time evolution happens from $+\infty$ to $-\infty$ due to which the fields at later times appear ``before" those at earlier times (of course, a more appropriate term to use is ``path-ordering" rather than ``time ordering"). The upper(lower) off-diagonal element is proportional to the negative(positive) frequency Wightman propagator. Path-ordering tells us that these are absolute-ordered correlations since the field on the forward branch must always appear ``before" the field on the backward branch. Explicit forms for the propagators of the short-wavelength fields are provided in Appendix A.
    
Let us expand $Z[\varphi_\pm]$ in powers of the coupling constant $\lambda$ which we will assume to be small for the perturbation theory to be valid. Although we will only calculate two-point correlation of the long-wavelength fields up to $\mathcal{O}(\lambda)$ we would like to keep terms up to $\mathcal{O}(\lambda^2)$ in the reduced generating functional to show how the various Feynman diagrams look in presence of the real source $J_i$ and how they differ with a similar work done previously in \cite{PerreaultLevasseur:2013kfq}. The expression for $Z[\varphi_\pm]$ up to $\mathcal{O}(\lambda^2)$ reads
        \begin{align}
            &Z[\varphi\pm]=\left\{1-\frac{i\lambda}{4}\varphi_i^2\eta_i\left(\frac{1}{i\eta_i}\frac{\delta}{\delta J_i}\right)^2+\frac{i^2\lambda^2}{2.(4)^2}\varphi_i^2\eta_i\varphi_j^2\eta_j\left(\frac{1}{i\eta_i}\frac{\delta}{\delta J_i}\right)^2\left(\frac{1}{i\eta_j}\frac{\delta}{\delta J_j}\right)^2\right\}\nonumber\\
            &\hspace{1.1cm}\left\{1-\frac{i\lambda}{3!}\varphi_i\eta_i\left(\frac{1}{i\eta_i}\frac{\delta}{\delta J_i}\right)^3+\frac{i^2\lambda^2}{2.(3!)^2} \varphi_i \eta_i \varphi_j \eta_j\left(\frac{1}{i\eta_i}\frac{\delta}{\delta J_i}\right)^3\left(\frac{1}{i\eta_j} \frac{\delta}{\delta J_j}\right)^3\right\}\nonumber\\
            &\hspace{1.1cm}\left\{1-\frac{i\lambda}{4!}\eta_i\left(\frac{1}{i\eta_i}\frac{\delta}{\delta J_i}\right)^4+\frac{i^2\lambda^2}{2.(4!)^2} \eta_i \eta_j\left(\frac{1}{i\eta_i}\frac{\delta}{\delta J_i}\right)^4\left(\frac{1}{i\eta_j} \frac{\delta}{\delta J_j}\right)^4\right\}Z_\text{f}[\varphi\pm;J_\pm].
        \end{align}
    From this we can obtain the expression for the influence functional up to an unimportant constant as follows:
    \begin{align}
        S_{\text{inf}}[\varphi_\pm]= \frac{1}{i}\text{ln} Z[\varphi_\pm],
    \end{align}
    where 
        \begin{align}
            &\frac{1}{i}\text{ln} Z[\varphi_\pm]=\frac{1}{i} \text{ln}Z_\text{f}[\varphi_\pm]+
            \frac{1}{i}\text{ln}\left[\left\{1-\frac{i\lambda}{4}\frac{\varphi_i^2\eta_i}{Z_\text{f}[\varphi_\pm;J_\pm]}\left(\frac{1}{i\eta_i}\frac{\delta}{\delta J_i}\right)^2
            -\frac{i\lambda}{3!}\frac{\varphi_i\eta_i}{Z_\text{f}[\varphi_\pm;J_\pm]}\left(\frac{1}{i\eta_i}\frac{\delta}{\delta J_i}\right)^3\right.\right. \nonumber\\
            &\hspace{1.5cm}\left.\left.-\frac{i\lambda}{4!}\frac{\eta_i}{Z_\text{f}[\varphi_\pm;J_\pm]}\left(\frac{1}{i\eta_i}\frac{\delta}{\delta J_i}\right)^4
            +\frac{i^2\lambda^2}{2.4^2}\frac{\varphi_i^2\eta_i\varphi_j^2\eta_j}{Z_\text{f}[\varphi_\pm;J_\pm]}\left(\frac{1}{i\eta_i}\frac{\delta}{\delta J_i}\right)^2\left(\frac{1}{i\eta_j}\frac{\delta}{\delta J_j}\right)^2\right.\right. \nonumber\\
            &\hspace{1.5cm}\left.\left.+\frac{i^2\lambda^2}{4.3!}\frac{\varphi_i^2\eta_i\varphi_j\eta_j}{Z_\text{f}[\varphi_\pm;J_\pm]}\left(\frac{1}{i\eta_i}\frac{\delta}{\delta J_i}\right)^2\left(\frac{1}{i\eta_j}\frac{\delta}{\delta J_j}\right)^3
            +\frac{i^2\lambda^2}{4.4!}\frac{\varphi_i^2\eta_i\eta_j}{Z_\text{f}[\varphi_\pm;J_\pm]}\left(\frac{1}{i\eta_i}\frac{\delta}{\delta J_i}\right)^2\left(\frac{1}{i\eta_j}\frac{\delta}{\delta J_j}\right)^4\right.\right. \nonumber\\
            &\hspace{1.5cm}\left.\left.+\frac{i^2\lambda^2}{2.(3!)^2}\frac{\varphi_i\eta_i\varphi_j\eta_j}{Z_\text{f}[\varphi_\pm;J_\pm]}\left(\frac{1}{i\eta_i}\frac{\delta}{\delta J_i}\right)^3\left(\frac{1}{i\eta_j}\frac{\delta}{\delta J_j}\right)^3
            +\frac{i^2\lambda^2}{3!4!}\frac{\varphi_i\eta_i\eta_j}{Z_\text{f}[\varphi_\pm;J_\pm]}\left(\frac{1}{i\eta_i}\frac{\delta}{\delta J_i}\right)^3\left(\frac{1}{i\eta_j}\frac{\delta}{\delta J_j}\right)^4\right.\right.\nonumber\\
            &\hspace{1.5cm}\left.\left.+\frac{i^2\lambda^2}{2.(4!)^2}\frac{\eta_i\eta_j}{Z_\text{f}[\varphi_\pm;J_\pm]}\left(\frac{1}{i\eta_i}\frac{\delta}{\delta J_i}\right)^4\left(\frac{1}{i\eta_j}\frac{\delta}{\delta J_j}\right)^4
            \right\}Z_\text{f}[\varphi_\pm;J_\pm]\right].
        \end{align}

   If we adopt the following Feynman rules:

   \begin{align}
        \hspace{1mm}
=(iG_{ij})(iJ_j\eta_j),
\end{align}
    
    we can write the influence functional in terms of the Feynman diagrams as follows:
\begin{align}\label{Sinfdiagram}
    S_{\text{inf}} &= \frac{1}{2i}\hspace{1mm}%
\right),
    \end{align}
   
    where a bar over indices indicates symmetrization over those indices for example: $\varphi^2_{\bar{i}}\eta_{\bar{i}}\varphi_{\bar{j}}\eta_{\bar{j}} = \frac{1}{2}(\varphi^2_{i}\eta_i\varphi_j\eta_j + \varphi_{i}\eta_i\varphi^2_j\eta_j)$.
    
The diagrammatic expressions for the various functional derivatives of $Z[\varphi_\pm,J_\pm]$ up to $\mathcal{O}(\lambda^2)$ used to obtain the influence functional can be found in Appendix \ref{Appendix B}. It can be seen that we have additional terms in (\ref{Sinfdiagram}) dependent on $J_i$ which have not been taken into account in \cite{PerreaultLevasseur:2013kfq}. As we mentioned earlier, the source $J_i = E_i\varphi_i$ has not been added by hand as a trick but arises naturally when we split the full field $\Phi$ according to (\ref{split}). As such, in general we are not allowed to put the source terms to zero and they may make a genuine contribution to the influence functional.

\section{Two-point correlation of long-wavelength fields at $\mathcal{O}(\lambda^0)$}\label{Sec3}

Our aim is to calculate the two-point correlation of the long-wavelength fields that takes into account the effect of the short-wavelength fields that have been integrated over in the path-integral. By keeping calculations up to zeroth order in $\lambda$ we merely repeat what has been done before for the sake of completeness in this section.
Let us begin our calculation with the first term of $S_{\text{inf}}$.
Writing the propagator as $G_{ij} = -iF_{ij}$ and expanding over the DeWitt indices, we have

\begin{align}\label{ft2}
    \frac{1}{2i}\hspace{1mm}\begin{tikzpicture}[baseline={([yshift=-.5ex]current bounding box.center)}, square/.style={regular polygon,regular polygon sides=4}]
    \begin{feynman}
    \node  at (0,0) [star,inner sep=0.2em,draw] (b) {};
    \node  at (1,0) [star,inner sep=0.2em,draw] (c) {};
    \diagram*{(b) -- [black] (c)};
    \end{feynman}
    \end{tikzpicture} = \frac{i}{2}\int d^4x \ a^3(t_x)\int d^4y \ a^3(t_y) \varphi_{I}(x) \ \eta_{I} \ \overrightarrow{E}_{I}(x) F_{IJ}(x,y) \overleftarrow{E}_{J}(y) \ \eta_{J} \ \varphi_{J}(y),
\end{align}

Keeping terms up to $\mathcal{O}(\lambda^0)$ and using the momentum-space expression for the propagators given in Appendix \ref{Appendix A} in Eq. (\ref{ft2}), we obtain

\begin{align}\label{ft01}
    \frac{i}{2}\int d^4x \ a^3(t_x)\int d^4y \ a^3(t_y) \varphi_{I}(x) \ \eta_{I} \ \overrightarrow{\Lambda}_{I}(x) \int \frac{d^3 k}{(2\pi)^3}W(k,t_x)W(k,t_y)e^{-i\mathbf{k}\cdot(\mathbf{x}-\mathbf{y})} f_{IJ}(\mathbf{k},t_x,t_y) \overleftarrow{\Lambda}_{J}(y) \ \eta_{J} \ \varphi_{J}(y) ,
\end{align}
where the definition of the matrix $f_{IJ}(\mathbf{k},t_x,t_y)$ has been provided in Appendix \ref{Appendix B}. If we take the window function in the free propagators of the short-wavelength fields to be the simple Heaviside theta function, then no undifferentiated window functions can be kept lying around in the expression above due to orthogonality. This can be explained as follows. Consider the following integral which is contained in the previous expression:
\begin{align} \label{window}
    \int d^3\mathbf{x} \ \varphi_+(x) W(k,t_x) e^{-i\mathbf{k}\cdot\mathbf{x}} &= \int d^3\mathbf{x} \int \frac{d^3 \mathbf{p}}{(2\pi)^3} \tilde\varphi_{\mathbf{p}}(t_x) e^{-i\mathbf{p}\cdot\mathbf{x}} W(k,t_x) e^{-i\mathbf{k}\cdot\mathbf{x}}\nonumber\\
    &=\tilde\varphi_{-\mathbf{k}}(t_x)W(k,t_x).
\end{align}
Since the long-wavelength modes are nonzero only for $|\mathbf{k}|<\sigma a(t_x) H$ and the window function being the Heaviside theta function has support only for $|\mathbf{k}|<\sigma a(t_x) H$ the expression above vanishes. When this argument is applied to Eq.(\ref{ft01}) we obtain
\begin{align}\label{ft02}
    \frac{i}{2}\int d^4x \ a^3(t_x)\int d^4y \ a^3(t_y) \varphi_I(x) \ \eta_{I} \int \frac{d^3 k}{(2\pi)^3}\overrightarrow{P}(k,t_x)e^{-i\mathbf{k}\cdot(\mathbf{x}-\mathbf{y})} f_{IJ}(\mathbf{k},t_x,t_y) \overleftarrow{P}(k,t_y) \ \eta_{J} \ \varphi_J(y),
\end{align}
where the operator $P(k,t)$ contains no undifferentiated window function:
\begin{align}\label{Pkt}
    P(k,t)= \ddot{W}(k,t) + 2\dot{W}(k,t)\frac{\partial}{\partial t} + 3H\dot{W}(k,t).
\end{align}
The influence functional has a real and an imaginary part. The contribution of the real part to the effective action of the long-wavelength fields is interpreted as dissipation (for details on this in the context of inflation see \cite{Morikawa:1989xz, Matarrese:2003ye, Tinwala:2024wod}). The interesting part of the influence functional is its imaginary contribution to the effective action. Indeed, one of the aims of this article is to investigate how the imaginary contribution of the influence functional can be interpreted. As such, we will focus exclusively on the imaginary part of the influence functional throughout this paper. If we shift a time derivative from $\ddot{W}(k,t)$ and expand the matrix products, then the imaginary part of Eq.(\ref{ft02}) reads
\begin{align}\label{ft01i}
    \frac{i}{2}\int d^4x \ a^3(t_x)\int d^4y \ a^3(t_y) (\Psi_+(x) - \Psi_-(x))\int \frac{d^3 k}{(2\pi)^3}A(\mathbf{k},t_x,t_y) (\Psi_+(y)-\Psi_-(y)).
\end{align}
The expressions for $\Psi(x)$ and $A(\mathbf{k},t_x,t_y)$ in (\ref{ft01i}) are given by
\begin{align}
    &\Psi_+(x) = \begin{pmatrix}
        \gamma_+(x) \\ \varphi_+(x),
    \end{pmatrix}\\
    &A(\mathbf{k},t_x,t_y) = \text{Re}\begin{pmatrix}
        \dot{W}(k,t_x)\dot{W}(k,t_y)\mathcal{A} & \dot{W}(k,t_x)\dot{W}(k,t_y)\mathcal{A}_y \\ \dot{W}(k,t_x)\dot{W}(k,t_y)\mathcal{A}_x & \dot{W}(k,t_x)\dot{W}(k,t_y)\mathcal{A}_{xy}
    \end{pmatrix},
\end{align}
where we have defined
\begin{align}
    &\gamma_{+/-}(x) = p_\phi(t_x)\varphi_+(x)-\dot\varphi(x)\\
    &\mathcal{A}(\mathbf{k},t_x,t_y) = e^{-i\mathbf{k}\cdot(\mathbf{x}-\mathbf{y})}\phi_\mathbf{k}(t_x)\phi^*_{\mathbf{k}}(t_y)\\
    &\mathcal{A}_x(\mathbf{k},t_x,t_y) = \frac{k}{a(t_x)H}e^{-i\mathbf{k}\cdot(\mathbf{x}-\mathbf{y})}\phi_\mathbf{k}(t_x)\phi^*_{\mathbf{k}}(t_y)\\
    &\mathcal{A}_y(\mathbf{k},t_x,t_y) = \frac{k}{a(t_y)H}e^{-i\mathbf{k}\cdot(\mathbf{x}-\mathbf{y})}\phi_\mathbf{k}(t_x)\phi^*_{\mathbf{k}}(t_y)\\
    &\mathcal{A}_{xy}(\mathbf{k},t_x,t_y) = \frac{k^2}{a(t_x)a(t_y)H^2}e^{-i\mathbf{k}\cdot(\mathbf{x}-\mathbf{y})}\phi_\mathbf{k}(t_x)\phi^*_{\mathbf{k}}(t_y).
\end{align}

The value of $p_\phi$ depends upon the mode functions. If we assume that the mode functions can be written in terms of Hankel functions then we have

\begin{align}
   &\phi_\mathbf{k}(t) = \frac{H\sqrt{\pi}}{2}\left(\frac{1}{aH}\right)^{3/2}H_\nu^{(1)}[k/(aH)],\\
   &\dot \phi_\mathbf{k}(t) = (p_\phi(t) + q_\phi[k/(aH)])\phi_\mathbf{k},
\end{align}
 where $|\mathbf{k}|=k$. For the case of minimally coupled massless scalars which is what we will be considering in this paper $p_\phi=0$ and $q_\phi = -H\left(i + \frac{1}{i+\frac{k}{aH}}\right)$. The form of the imaginary influence functional Eq.(\ref{ft01i}) is such that it can be given an interpretation in terms of a classical random variable. This is achieved by making use of the Hubbard-Stratonovich (HS) transformation:
 \begin{align}
     \exp{\left(-\frac{1}{2}x_iA_{ij}x_j\right)} = \frac{1}{N}\prod_{i}\left(\int_{-\infty}^{\infty} d\xi_i\right)\exp{\left(-\frac{1}{2}\xi_i A^{-1}_{ij}\xi_j\right)}\exp{(i\xi_ix_i)}\label{hs1},
 \end{align}
 where $N = \prod_{i}\left(\int_{-\infty}^{\infty} d\xi_i\right)\exp{\left(-\frac{1}{2}\xi_i A^{-1}_{ij}\xi_j\right)}$.
On applying this transformation to $\exp{(iS_{\text{inf}})}$ where $S_{\text{inf}}$ is given by Eq.(\ref{ft01i}) we obtain
\begin{align}\label{Sinf0}
    \exp{(i S_\text{inf}}) =  &= \frac{1}{N}\int \mathscr{D}\xi_1\int\mathscr{D}\xi_2 \exp{\left(-\frac{1}{2}\int d^4x \ a^3(t_x)\int d^4y \ a^3(t_y) \xi_I(x)\left\{\int\frac{d^3 k}{(2\pi)^3}A(k,t_x,t_y)\right\}^{-1}_{IJ}\xi_J(y)\right)}\nonumber\\
    &\hspace{3cm}\times\exp\left(i\int d^4x  \ a^3(t_x)\Big\{\xi_1(x)\gamma_q(x) + \xi_2(x)\varphi_q(x)\Big\}\right),
\end{align}
where we have made use of the Keldysh basis

\begin{align}
    &\varphi_q(x) = \varphi_+(x)-\varphi_{-}(x)\\
    &\gamma_q(x) = \gamma_{+}(x)-\gamma_{-}(x)\\
    &\varphi_c=(\varphi_++\varphi_-)/2\\
    &\gamma_c = (\gamma_++\gamma_-)/2.
\end{align}

Through the application of the HS transformation we have re-introduced the path integral now over the classical fields $\xi_1$ and $\xi_2$ with an additional gaussian term given by

\begin{align}\label{pb0}
    \mathscr{P}[\xi_1,\xi_2]=\frac{1}{N}\exp{\left(-\frac{1}{2}\int d^4x \ a^3(t_x)\int d^4y \ a^3(t_y) \xi_I(x)\left\{\int\frac{d^3 k}{(2\pi)^3}A(k,t_x,t_y)\right\}^{-1}_{IJ}\xi_J(y)\right)}.
\end{align}

We have also obtained a new action defined by

\begin{align}
    \Delta S[\varphi_\pm] = \int d^4x  \ a^3(t_x)\Big\{\xi_1(x)\gamma_q(x) + \xi_2(x)\varphi_q(x)\Big\}.
\end{align}

The interpretation of (\ref{Sinf0}) is that $e^{iS_{\text{inf}}}$ is equal to $e^{i\Delta S}$ averaged over all configurations of the fields $\xi_1$ and $\xi_2$ which are now treated as random and following the normalized probability distribution function given by (\ref{pb0}). The effective action for the long-wavelength fields can be written as
\begin{align}
    &\exp{(i\Gamma)} = \exp{\left[i\left(S[\varphi^+]-S[\varphi^-]+S_{\text{inf}}\right)\right]} = \int \mathscr{D}\xi_1\int\mathscr{D}\xi_2 \ \mathscr{P}[\xi_1,\xi_2]\exp{[i S_{\text{eff}}]};
\end{align}
where we have defined an effective action $S_\text{eff}$ without the average over the random fields as

\begin{align}\label{Seff0}
    S_{\text{eff}} = S_{+}[\varphi_+] - S_{-}[\varphi_-] +\int d^4x  \ a^3(t_x)(\xi_1(x)\gamma_q(x) + \xi_2(x)\varphi_q(x)).
\end{align}

The equations of motion follow from
\begin{align}
    \frac{\delta S_{\text{eff}}}{\delta\varphi_q(x)}\Bigg|_{\varphi_q=0} = 0 = \ddot\varphi(x) + 3H\dot\varphi(x) -\frac{\nabla^2}{a^2}\varphi(x) - 3H\xi_1(x) - \dot\xi_1(x) + \xi_2(x)\label{eqmotion1}.
\end{align}
If slow roll condition ($\ddot\varphi<<3H\dot\varphi$) is assumed and $\sigma<<1$ is used then the equation for the long-wavelength fields $(|\mathbf{k}|<<aH)$ becomes
\begin{align}\label{eqmotion0}
    \dot\varphi = \xi_1(t).
\end{align}
The limit $\sigma<<1$ is required so that the correlation functions of the long-wavelength fields calculated using the stochastic formalism match with those calculated using quantum field theory \cite{Grain:2017dqa}\footnote{Note that the $\sigma=0$ limit is employed in \cite{Tsamis:2005hd} in a different manner without introducing $\sigma$}. The reason behind neglecting $\dot\xi_1$ and $\xi_2$ is that the correlations involving these variables (like $\langle\dot\xi_1(t)\xi_1(t')\rangle$) contains extra factors of $k$ which are replaced either by $\sigma a(t)H$ or $\sigma a(t')H$ owing to the Dirac delta functions arising from the derivatives of the window functions. This means that such terms contain extra factors of $\sigma$ which can be dropped on the account of taking $\sigma<<1$. We refer the readers to Appendix \ref{Appendix C} for a detailed explanation on this subject. On solving (\ref{eqmotion0}) we obtain the following standard result for the two-point correlation of the long-wavelength fields \cite{Morikawa:1989xz, Matarrese:2003ye}:

\begin{align}
    \langle\varphi(t)\varphi(t')\rangle = \frac{H^3}{4\pi^2}(t-t_i) = \frac{H^2}{4\pi^2}\log{\left(\frac{a(t)}{a(t_i)}\right)},
\end{align}

where we assumed $t<t'$ and $t_i$ to be the initial time.

\section{Computation of the two-point correlation of long-wavelength fields up to $\mathcal{O}(\lambda)$}\label{Sec4}

Let us now retain the $\lambda$ dependent terms in the first diagram of $S_{\text{inf}}$ Eq.(\ref{ft2}). We wish to calculate its contribution to the two-point correlation of the long-wavelength fields. The extra terms in Eq.(\ref{ft2}) that we need to calculate are

\begin{align}\label{extra}
    -\frac{i}{2}\eta_i\varphi_i\overrightarrow{\Lambda}_iF_{ij}\left(\frac{\lambda\varphi_j^3}{3!}\right)\eta_j -\frac{i}{2}\eta_i\left(\frac{\lambda\varphi_i^3}{3!}\right)F_{ij}\overleftarrow{\Lambda}_j\varphi_j\eta_j + \frac{i}{2}\eta_i\left(\frac{\lambda\varphi_i^3}{3!}\right)F_{ij}\left(\frac{\lambda\varphi_j^3}{3!}\right)\eta_j.
\end{align}

Using the momentum-space expression of the propagator in the first term of the above expression, we obtain the following expression for the imaginary part:

\begin{align}
    -\frac{i\lambda}{2}\eta_i\varphi_i\overrightarrow{\Lambda}_iF_{ij}\left(\frac{\varphi_j^3}{3!}\right)\eta_j &= -\frac{i\lambda}{2}\int d^4x \ a^3(t_x)\int d^4y \ a^3(t_y) \eta_I\varphi_I(x)\nonumber\\
    &\hspace{1cm}\times\left\{\overrightarrow{\Lambda}_I(x)\int \frac{d^3\mathbf{k}}{(2\pi)^3} W(k,t_x) W(k,t_y) e^{-i\mathbf{k}\cdot(\mathbf{x}-\mathbf{y})} f^R_{IJ}(k,t_x,t_y)\right\} \ \frac{\varphi_J^3(y)}{3!}\eta_J\label{extra1a} \\
    &=-\frac{i\lambda}{2}\int d^4x \ a^3(t_x)\int d^4y \ a^3(t_y)\eta_I\varphi_I(x)\nonumber\\
    &\hspace{1cm}\times\left\{\int \frac{d^3\mathbf{k}}{(2\pi)^3} \overrightarrow{P}(k,t_x) e^{-i\mathbf{k}\cdot(\mathbf{x}-\mathbf{y})} f^R_{IJ}(k,t_x,t_y) W(k,t_y)\right\} \ \frac{\varphi_J^3(y)}{3!}\eta_J\label{extra1b},
\end{align}

where the operator $P(k,t_x)$ appears above for the same reasons as given in the previous section. The integral over $\mathbf{y}$ requires some attention and further analysis of it is done in the following subsection.

\subsection{Orthogonality analysis}\label{4A}

The expression (\ref{extra1a}) is linear in $\varphi_I(x)$ but cubic in $\varphi_J(y)$. Because it is linear in $\varphi_I(x)$, the same orthogonality argument used in the previous section was applied here in going from (\ref{extra1a}) to (\ref{extra1b}). The effect of this is that the operator $\Lambda$ turns into $P(k,t_x)$. However, a careful analysis is required to check if the same orthogonality argument can be used for the integral over $y$ since it involves cubic power of the long-wavelength field. For simplicity, let us first consider the following integral:
\begin{align}\label{ortho2a}
    \int d^3y \ \varphi^N_J(y) W(k,t_y) e^{i\mathbf{k}\cdot\mathbf{y}}, \hspace{2cm}\text{with}\hspace{2mm}N=2.
\end{align}
Later we will generalize this for any arbitrary positive integer power of $\varphi_J(x)$.
Fourier transforming $\varphi^2_J(y)$ we find
\begin{align}
     \int d^3y \ \varphi^2_J(y) W(k,t_y) e^{i\mathbf{k}\cdot\mathbf{y}} &= \int d^3y \int \frac{d^3\mathbf{p}_1}{(2\pi)^3}\int \frac{d^3\mathbf{p}_2}{(2\pi)^3} e^{-i\mathbf{p}_1\cdot\mathbf{y}}e^{-i\mathbf{p}_2\cdot\mathbf{y}}\tilde\varphi_{\mathbf{p}_1}\tilde\varphi_{\mathbf{p}_2}\times W(k,t_y)e^{i\mathbf{k}\cdot\mathbf{y}} \nonumber\\
     &=\int \frac{d^3\mathbf{p}}{(2\pi)^3}\tilde\varphi_{\mathbf{k}-\mathbf{p}}\tilde\varphi_\mathbf{p} W(k,t_y)\label{ortho2b}.
\end{align}
The definitions of the long-wavelength fields and the window function dictate the following inequalities to be satisfied by the various momenta.
\begin{align}
    |\mathbf{k}-\mathbf{p}|<\sigma a(t_y)H, \hspace{5mm}|\mathbf{p}|<\sigma a(t_y) H, \hspace{5mm}|\mathbf{k}|>\sigma a(t_y)H.
\end{align}
The first inequality can be written as
\begin{align}
    k^2 + p^2 -2kp\cos\theta < \sigma^2 a^2H^2.
\end{align}
It becomes possible now to allow $k$ to be greater than $\sigma a H$ and yet have a non-zero value of the expression (\ref{ortho2b}). The maximum value that $k$ can assume can be found by solving the quadratic equation:
\begin{align}
     k^2 + p^2 -2kp\cos\theta = \sigma^2 a^2H^2,
\end{align}
whose solutions are
\begin{align}
    k_\pm = \frac{2p\cos\theta \pm{4p^2\cos^2\theta - 4(p^2-\sigma^2a^2H^2)}^{1/2}}{2} .
\end{align}
To satisfy $(k-k_+)(k-k_-)<0$, $k$ must be in the range $k_-<k<k_+$. This range can be enlarged to maximum by taking $p=\sigma a H$ and $\theta=0$ which results in $\sigma a H<k<2\sigma a H$. This result can be generalized to any arbitrary positive integer power $N$ of $\varphi_J(y)$ in Eq.(\ref{ortho2a}). Repeating this exercise, we find the following inequalities:
\begin{align}
    &k^2 + \sum_{i=1}^{N-1}p_i^2 - 2k\sum_{i=1}^{N-1}p_i\cos\theta_i + 2\sum_{i,j=1, \ i<j}^{N-1}p_ip_j\cos\theta_{ij}<\sigma^2a^2H^2\\
    &p_i<\sigma a H\\
    &k>\sigma a H.
\end{align}
Taking the internal momenta $p_i=\sigma a H$, the angles between $\mathbf{k}$ and the internal momenta, $\theta_i=0$, and the angles between the internal momenta $\theta_{ij}=0$ the range of $k$ can be maximized to
\begin{align}
    \sigma a H<k< N\sigma a H.
\end{align}
For any $k>N\sigma a H$, the integral Eq.(\ref{ortho2a}) vanishes regardless of what values $p_i$ and the various angles can take. We can see that for $N=1$ Eq.(\ref{ortho2a}) vanishes, which is the argument we used earlier. For the case of quartic interaction, we have $N=3$ and so the maximum allowed range for $k$ for which Eq.(\ref{ortho2a}) could be nonzero is 
\begin{align}
    \sigma a H<k<3\sigma a H.
\end{align}
We observe that there exists a range of $k$ for which Eq. (\ref{ortho2a}) does not vanish even when $k> \sigma a H$. As such the modes cannot be considered strictly orthogonal, and the contributions from these terms must also be included. The detailed computation incorporating these additional contributions has been performed in the next subsection.\\
Crucially, this feature is not an artifact of the specific window function adopted here, but would persist for a broad class of window functions. As seen in (\ref{window}), the contribution vanished only because the Heaviside theta function enforced a complete separation between long- and short-wavelength fields, allowing them to be treated as strictly orthogonal. However, with any other choice of window function, such a separation does not hold, and the long- and short-wavelength fields cannot, in general, be assumed orthogonal.  By the same reasoning, one inevitably encounters additional contributions in the analysis for $N>1$ cases as well, analogous to what we have obtained here for the Heaviside theta function. 
\subsection{Complete imaginary part of the first diagram of $S_{\text{inf}}$}\label{4B}
We now proceed to calculate Eq.(\ref{extra}). A rigorous calculation would require integration over internal momenta and angles. Here, however, we keep it simple by choosing the maximum allowed range for $k$. In this way, the result will only be approximate, but it serves the purpose of estimating the effect of the extra terms not previously considered in the literature. In fact, the result we will obtain would be an upper bound, since we are using the maximum allowed range of $k$, taking specific values for the internal momenta and the angles. The full calculation will necessarily be less than what we will obtain here. Keeping the integral limits on $k$ implicit for time being, expanding over DeWitt indices and shifting one time derivative from $\ddot W$ in operator $P$, the imaginary part of the extra terms in Eq.(\ref{extra}) can be computed further as follows:

\begin{align}\label{extra2}
    &-\frac{i}{2}\eta_i\varphi_i\overrightarrow{\Lambda}_iF_{ij}\left(\frac{\lambda\varphi_j^3}{3!}\right)\eta_j -\frac{i}{2}\eta_i\left(\frac{\lambda\varphi_i^3}{3!}\right)F_{ij}\overleftarrow{\Lambda}_j\varphi_j\eta_j + \frac{i}{2}\eta_i\left(\frac{\lambda\varphi_i^3}{3!}\right)F_{ij}\left(\frac{\lambda\varphi_j^3}{3!}\right)\eta_j\nonumber\\
    =&\frac{i}{2}\int d^4x \ a^3(t_x)\int d^4y \ a^3(t_y) (\Xi^T_+(x)-\Xi^T_-(x))\int \frac{d^3\mathbf{k}}{(2\pi)^3} B(\mathbf{k},t_x,t_y) (\Xi_+(y)-\Xi_-(y)),
\end{align}
where we have defined
\begin{align}
    &\Xi_\pm = \begin{pmatrix}
        \gamma_\pm \\ \varphi_\pm \\ -\frac{\lambda\varphi_\pm}{3!}
    \end{pmatrix},\\
    &B(\mathbf{k},t_x,t_y) = \text{Re}\begin{pmatrix}
        0 & 0 & \dot W(k,t_x)W(k,t_y)\mathcal{A}\\
        0 & 0 & \dot W(k,t_x)W(k,t_y)\mathcal{A}_x\\
        W(k,t_x)\dot W(k,t_y)\mathcal{A} & W(k,t_x)\dot W(k,t_y)\mathcal{A}_y & W(k,t_x)W(k,t_y)\mathcal{A}
    \end{pmatrix}.
\end{align}
Combining this result with the $\mathcal{O}(\lambda^0)$ result in (\ref{ft01i}) we obtain the following expression for the imaginary part of the first diagram in $S_\text{inf}$ (\ref{ft2}):
\begin{align}
    i\text{Im}\left[ \frac{1}{2i}\hspace{1mm}\begin{tikzpicture}[baseline={([yshift=-.5ex]current bounding box.center)}, square/.style={regular polygon,regular polygon sides=4}]
    \begin{feynman}
    \node  at (0,0) [star,inner sep=0.2em,draw] (b) {};
    \node  at (1,0) [star,inner sep=0.2em,draw] (c) {};
    \diagram*{(b) -- [black] (c)};
    \end{feynman}
    \end{tikzpicture}\right] = \frac{i}{2}\int d^4x \ a^3(t_x)\int d^4y \ a^3(t_y) (\Xi^T_+(x)-\Xi^T_-(x))\int \frac{d^3\mathbf{k}}{(2\pi)^3} A_\lambda(\mathbf{k},t_x,t_y) (\Xi_+(y)-\Xi_-(y))\label{extrai1},
\end{align}
where the imaginary part of $A$ is defined as $\text{Im}A = \dfrac{A-A^*}{2i}$ and $A_\lambda$ is given by
\begin{align}
    A_\lambda(\mathbf{k},t_x,t_y) = \text{Re}\begin{pmatrix}
        \dot{W}(k,t_x)\dot{W}(k,t_y)\mathcal{A} & \dot{W}(k,t_x)\dot{W}(k,t_y)\mathcal{A}_y & \dot W(k,t_x)W(k,t_y)\mathcal{A}\\
        \dot{W}(k,t_x)\dot{W}(k,t_y)\mathcal{A}_x & \dot{W}(k,t_x)\dot{W}(k,t_y)\mathcal{A}_{xy} & \dot W(k,t_x)W(k,t_y)\mathcal{A}_x\\
        W(k,t_x)\dot W(k,t_y)\mathcal{A} & W(k,t_x)\dot W(k,t_y)\mathcal{A}_y & W(k,t_x)W(k,t_y)\mathcal{A}.
    \end{pmatrix}.
\end{align}
The integration limits of $\mathbf{k}$ in the above expression for matrix elements in the third row and the third column of $A_\lambda$ will be carefully worked out in the next section due to the result of the orthogonality analysis carried out earlier. 
\subsection{Noise correlations}\label{4C}
The complete expression for the imaginary part of the first diagram in $S_\text{inf}$ given by (\ref{extrai1}) has a form such that the Hubbard-Stratonovich (HS) transformation could be applied. The result is as follows:
\begin{align}
      \exp{(iS_\text{inf}}) =  &= \frac{1}{N}\left(\prod_{I=1}^3\int \mathscr{D}\xi_I\right)\exp{\left(-\frac{1}{2}\int d^4x \ a^3(t_x)\int d^4y \ a^3(t_y) \xi_I(x)\left\{\int\frac{d^3 k}{(2\pi)^3}A_\lambda(\mathbf{k},t_x,t_y)\right\}^{-1}_{IJ}\xi_J(y)\right)}\nonumber\\
    &\hspace{3cm}\times\exp{(i\int d^4x  \ a^3(t_x)\left\{\xi_1(x)\gamma_q(x) + \xi_2(x)\varphi_q(x) -\frac{\lambda}{3!}(\varphi_+^3(x)-\varphi_-^3(x))\xi_3(x)\right\}}\label{hs2}.
\end{align}
We have an additional noise variable $\xi_3$ this time corresponding to the quantity $-\frac{\lambda}{3!}(\varphi_+^3(x)-\varphi_-^3(x))$.
Let us calculate the additional correlations that involve $\xi_3$. In what follows, we will ignore all those matrix elements that involve either $\mathcal{A}_x$, $\mathcal{A}_y$ or $\mathcal{A}_{xy}$ because of reasons explained in Appendix \ref{Appendix C}. In addition, since we wish to calculate the two-point correlation of long-wavelength fields only up to $\mathcal{O}(\lambda)$ we can ignore the $\langle\xi_3(x)\xi_3(y)\rangle$ correlation. This will become clear when we solve the equation of motion and calculate the two-point correlation of the long-wavelength field. As such, the only relevant correlation to be calculated is $\langle\xi_1(x)\xi_3(y)\rangle$ which is given by
\begin{align}\label{xi1xi30}
    \langle\xi_1(x)\xi_3(y)\rangle = \int \frac{d^3\mathbf{k}}{(2\pi)^3}(A_\lambda(\mathbf{k},t_x,t_y))_{13}.
\end{align}
The expression for $\text{Re} \ \mathcal{A}(\mathbf{k},t_x,t_y)$ is obtained using the mode functions for the minimally coupled massless scalar associated with Bunch-Davies vacuum. Since we are interested in $\langle\varphi(x)\varphi(y)\rangle$ for $|\mathbf{x}-\mathbf{y}|=0$, the expression for $\text{Re} \ \mathcal{A}(\mathbf{k},t_x,t_y)$ reads
\begin{align}
    \text{Re} \ \mathcal{A}(\mathbf{k},t_x,t_y) = \frac{H^2}{2k^3}\left\{\left(1+\frac{k^2}{H^2a_xa_y}\right)\cos\left[\frac{k}{H}\left(\frac{1}{a_x}-\frac{1}{a_y}\right)\right] + \frac{k}{H}\left(\frac{1}{a_x}-\frac{1}{a_y}\right)\sin\left[ \frac{k}{H}\left(\frac{1}{a_x}-\frac{1}{a_y}\right)\right]\right\}.
\end{align}

The matrix element $(A_\lambda(\mathbf{k},t_x,t_y))_{13}$ then reads
\begin{align}
   (A_\lambda(k,t_x,t_y))_{13}= \dot W(k,t_x)W(k,t_y)\text{Re} \ \mathcal{A}(k, t_x, t_y) = -kH \ \delta(k-\sigma a_xH) \ \theta\left(\frac{k}{\sigma a_yH}-1\right)\text{Re} \ \mathcal{A}(\mathbf{k},t_x,t_y),
\end{align}
where $k = |\mathbf{k}|$. The orthogonality analysis performed in the previous subsection gives a maximum upper bound of $N\sigma a_yH$ on $k$ which can be included in the above expression by using the Heaviside theta function as follows:
\begin{align}\label{A13}
    (A_\lambda(\mathbf{k},t_x,t_y))_{13} = -kH \ \delta(k-\sigma a_xH) \ \theta\left(\frac{k}{\sigma a_yH}-1\right) \ \theta\left(\frac{N\sigma a_yH}{k}-1\right)\text{Re} \ \mathcal{A}(\mathbf{k},t_x,t_y).
\end{align}
It is clear from $\delta$ function that $(A_\lambda(\mathbf{k},t_x,t_y))_{13}$ will be non-zero only if $\sigma a_y H<\sigma a_xH<N\sigma a_y H$. Since we take the limit $\sigma<<1$ in the end, we can expand $\text{Re} \ \mathcal{A}(k=\sigma a_x H,t_x, t_y)$ to least order in $\sigma$,
\begin{align}
    \text{Re} \ \mathcal{A}(k=\sigma a_x H,t_x, t_y) \approx \frac{H^2}{2k^3}\Bigg|_{k=\sigma a_x H}.
\end{align}
Using this result in (\ref{A13}) and integrating over $\mathbf{k}$, we obtain the following expression for the correlation in (\ref{xi1xi30}):
\begin{align}
  \langle\xi_1(t_x)\xi_3(t_y)\rangle = \int \frac{d^3\mathbf{k}}{(2\pi)^3}(A_\lambda(\mathbf{k},t_x,t_y))_{13} &= \frac{H^3}{4\pi^2} \ \theta\left(\frac{a_x}{a_y}-1\right) \ \theta\left(\frac{N a_y}{a_x}-1\right)\nonumber\\
  &=\frac{H^3}{4\pi^2} \ \theta\left(t_x-t_y\right) \ \theta\left(\beta + t_y-t_x\right)\label{xi1xi3},
\end{align}
where we have defined $\beta = \frac{1}{H}\ln N$. An important aspect of the result in (\ref{xi1xi3}) is that it is independent of $\sigma$ which means it will survive in the end when we take the limit $\sigma<<1$.

\subsection{Two-point correlation of long-wavelength fields at $\mathcal{O}(\lambda)$}\label{4D}
The transformation given in (\ref{hs2}) yields the following expression for the effective action:
\begin{align}
    S_{\text{eff}} = S_+[\varphi_+] - S_-[\varphi_-] + \int d^4x \ a^3_x\left\{\xi_1(x)\gamma_q(x) + \xi_2(x)\varphi_q(x) - \frac{\lambda}{3!}(\varphi_+^3(x)-\varphi_-^3(x))\xi_3(x)\right\}.
\end{align}
The equations of motion that follow after assuming slow roll conditions and ignoring $\xi_2$ and $\dot\xi_1$ (in the limit $\sigma<<1$) reads
\begin{align}
    3H\dot\varphi + \frac{\lambda\varphi^3}{6} - 3H\xi_1 + \frac{\lambda\varphi^2}{2}\xi_3 = 0 .
\end{align}
Assuming a perturbative expansion $\varphi = \varphi_0 + \lambda\varphi_1 + \mathcal{O}(\lambda^2)$ we can solve the equation for the long-wavelength fields up to $\mathcal{O}(\lambda)$ to obtain
\begin{align}
    &\varphi_0(T) = \int_{t_i}^{T} dt \ \xi_1(t),\\
    &\varphi_1(T) = -\frac{1}{3H}\int_{t_i}^{T}dt\left\{\frac{1}{6}\left(\int_{t_i}^{t} dt' \xi_1(t)\right)^3 + \frac{1}{2}\left(\int_{t_i}^{t}dt'\xi_1(t')\right)^2\xi_3(t)\right\}.
\end{align}
The two-point correlation at $\mathcal{O}(\lambda)$ is given by
\begin{align}\label{2ptl1}
    \lambda\langle\varphi_0(T)\varphi_1(T')\rangle + \lambda\langle\varphi_0(T')\varphi_1(T)\rangle.
\end{align}
We simply require to calculate the first term in the above expression since the second term can be obtained by swapping $T\leftrightarrow T'$. The first term reads
\begin{align}
    \lambda\langle\varphi_0(T)\varphi_1(T')\rangle &= -\frac{\lambda}{18H}\int_{t_i}^Tdt\int_{t_i}^{T'}dt'\int_{t_i}^{t'} dt_1\int_{t_i}^{t'} dt_2\int_{t_i}^{t'}dt_3\langle\xi_1(t)\xi_1(t_1)\xi_1(t_2)\xi_1(t_3)\rangle\nonumber\\
    &\hspace{2cm}-\frac{\lambda}{6H}\int_{t_i}^Tdt\int_{t_i}^{T'}dt'\int_{t_i}^{t'} dt_1\int_{t_i}^{t'} dt_2\langle\xi_1(t)\xi_1(t_1)\xi_1(t_2)\xi_3(t')\rangle\label{phi0phi1a}.
\end{align}
Let us calculate the first term of the above expression. Since $\xi_I$ follows a Gaussian probability distribution, the four-point noise correlator can be written in terms of the product of two-point correlators using Wick's theorem, which implies
\begin{align}
    \langle\xi_1(t)\xi_1(t_1)\xi_1(t_2)\xi_1(t_3)\rangle &= \langle\xi_1(t)\xi_1(t_1)\rangle\langle\xi_1(t_2)\xi_1(t_3)\rangle + \langle\xi_1(t)\xi_1(t_2)\rangle\langle\xi_1(t_1)\xi_1(t_3)\rangle + \langle\xi_1(t)\xi_1(t_3)\rangle\langle\xi_1(t_1)\xi_1(t_2)\rangle\nonumber\\
    &=3\langle\xi_1(t)\xi_1(t_1)\rangle\langle\xi_1(t_2)\xi_1(t_3)\rangle\label{4xi1},
\end{align}
where we have used the fact that the first term in (\ref{phi0phi1a}) is invariant under permutations of $t_1$, $t_2$ and $t_3$. Using the expression of $\langle\xi_1(t)\xi_1(t')\rangle$ from Appendix \ref{Appendix C} we obtain 

\begin{align}
     &-\frac{\lambda}{18H}\int_{t_i}^Tdt\int_{t_i}^{T'}dt'\int_{t_i}^{t'} dt_1\int_{t_i}^{t'} dt_2\int_{t_i}^{t'}dt_3\langle\xi_1(t)\xi_1(t_1)\xi_1(t_2)\xi_1(t_3)\rangle \nonumber\\&\hspace{5cm}=   -\frac{3\lambda}{18H}\left(\frac{H^3}{4\pi^2}\right)^2\int_{t_i}^Tdt\int_{t_i}^{T'}dt'\int_{t_i}^{t'} dt_1\int_{t_i}^{t'} dt_2\int_{t_i}^{t'}dt_3 \delta(t-t_1)\delta(t_2-t_3).
\end{align}

We now make use of the identity

\begin{align}\label{id1}
   \int_{a}^b dx\int_a^c dy \ \delta(x-y)=\begin{cases}
   \theta(b-c)\int_a^c dy + \theta(c-b)\int_a^b dx, & \text{if $b\neq c$}\\
   \int_a^b dx, & \text{if $b=c$}
  \end{cases}\hspace{0.5cm},
\end{align}

to obtain

\begin{align}
    &-\frac{3\lambda}{18H}\left(\frac{H^3}{4\pi^2}\right)^2\int_{t_i}^{T'} dt'\int_{t_i}^T dt\int_{t_i}^{t'} dt_1\delta(t-t_1)\int_{t_i}^{t'} dt_2\int_{t_i}^{t'}dt_3 \delta(t_2-t_3)\nonumber\\ &= -\frac{3\lambda}{18H}\left(\frac{H^3}{4\pi^2}\right)^2\int_{t_i}^{T'}dt'\left\{\theta(T-t')\int_{t_i}^{t'} dt_1 + \theta(t'-T)\int_{t_i}^Tdt\right\}\int_{t_i}^{t'} dt_2\nonumber\\
    &=-\frac{3\lambda}{18H}\left(\frac{H^3}{4\pi^2}\right)^2\left[\int_{t_i}^{T'} dt'\theta(T-t')(t'-t_i)^2 + (T-t_i)\int_{t_i}^{T'} dt'\theta(t'-T)(t'-t_i)\right]\nonumber\\
    &=-\frac{3\lambda}{18H}\left(\frac{H^3}{4\pi^2}\right)^2\left[\theta(T-T')\int_{t_i}^{T'} dt'(t'-t_i)^2 + \theta(T'-T)\int_{t_i}^T dt' (t'-t_i)^2 + \theta(T-T')\times 0 \right.\nonumber\\
    &\hspace{3cm}\left.+ \theta(T'-T)(T-t_i)\int_{T}^{T'}dt'(t'-t_i)\right]\nonumber\\
    &=-\frac{3\lambda}{18H}\left(\frac{H^3}{4\pi^2}\right)^2\left[\theta(T-T')\frac{(T'-t_i)^3}{3} + \theta(T'-T)\left\{\frac{(T-t_i)^3}{3} + (T-t_i)\left(\frac{(T'-t_i)^2}{2}-\frac{(T-t_i)^2}{2}\right)\right\}\right]\nonumber\\
    &=-\frac{3\lambda}{18H}\left(\frac{H^3}{4\pi^2}\right)^2\left[\theta(T-T')\frac{(T'-t_i)^3}{3} + \theta(T'-T)\left\{\frac{(T-t_i)(T'-t_i)^2}{2} - \frac{(T-t_i)^3}{6}\right\}\right].
\end{align}

Thus, we obtain the following expression for the first term of (\ref{phi0phi1a}):

\begin{align}
    &-\frac{\lambda}{18H}\int_{t_i}^Tdt\int_{t_i}^{T'}dt'\int_{t_i}^{t'} dt_1\int_{t_i}^{t'} dt_2\int_{t_i}^{t'}dt_3\langle\xi_1(t)\xi_1(t_1)\xi_1(t_2)\xi_1(t_3)\rangle\nonumber\\
    &=-\frac{3\lambda}{18H}\left(\frac{H^3}{4\pi^2}\right)^2\left[\theta(T-T')\frac{(T'-t_i)^3}{3} + \theta(T'-T)\left\{\frac{(T-t_i)(T'-t_i)^2}{2} - \frac{(T-t_i)^3}{6}\right\}\right]\label{res1}.
\end{align}

Let us compute the second term of (\ref{phi0phi1a}). Using Wick's Theorem, we obtain

\begin{align}\label{phi0phi11b}
    &-\frac{\lambda}{6H}\int_{t_i}^Tdt\int_{t_i}^{T'}dt'\int_{t_i}^{t'} dt_1\int_{t_i}^{t'} dt_2\langle\xi_1(t)\xi_1(t_1)\xi_1(t_2)\xi_3(t')\rangle \nonumber\\
    &= -\frac{\lambda}{6H}\int_{t_i}^Tdt\int_{t_i}^{T'}dt'\int_{t_i}^{t'} dt_1\int_{t_i}^{t'} dt_2\left\{2\langle\xi_1(t)\xi_1(t_1)\rangle\langle\xi_1(t_2)\xi_3(t')\rangle + \langle\xi_1(t)\xi_3(t')\rangle\langle\xi_1(t_1)\xi_1(t_2)\rangle\right\}.
\end{align}

The first term of (\ref{phi0phi11b}) vanishes

\begin{align}
   -\frac{2\lambda}{6H}\left(\frac{H^3}{4\pi^2}\right)^2\int_{t_i}^Tdt\int_{t_i}^{T'}dt'\int_{t_i}^{t'}dt_1\int_{t_i}^{t'}dt_2 \ \delta(t-t_1) \ \theta(t_2-t') \ \theta(\beta+t'-t_2) = 0.
\end{align}

because $\theta(t_2-t')$ enforces $t_2>t'$ while $t'$ itself serves as the upper limit of integration for $t_2$. The second term of (\ref{phi0phi11b}) reads

\begin{align}
   -\frac{\lambda}{6H}\left(\frac{H^3}{4\pi^2}\right)^2\int_{t_i}^Tdt\int_{t_i}^{T'}dt'\int_{t_i}^{t'}dt_1\int_{t_i}^{t'}dt_2 \ \delta(t_1-t_2) \ \theta(t-t') \ \theta(\beta+t'-t).
   \end{align}

We use the identity (\ref{id1}) and the following one:

\begin{align}
    \int_a^bdx\int_a^cdy \ \theta(x-y) \ \theta(\beta + y-x) &= \left\{\int_y^{\beta 
 + y}dx\int_a^{y_p}dy + \int_y^bdx\int_{y_p}^cdy\right\}\theta(\beta + c-b) \ \theta(b-c) \nonumber\\
    &+ \int_y^{\beta + y}dx\int_a^c dy \ \theta(b-(\beta + c)) \ \theta(b-c)  \nonumber\\
    &+ \left\{\int_y^{\beta + y}dx\int_a^{y_p}dy + \int_y^bdx\int_{y_p}^bdy\right\} \ \theta(c-b),
\end{align}

where $y_p = b-\beta $, to obtain

\begin{align}\label{res2}
   &-\frac{\lambda}{6H}\left(\frac{H^3}{4\pi^2}\right)^2\int_{t_i}^Tdt\int_{t_i}^{T'}dt'\int_{t_i}^{t'}dt_1\int_{t_i}^{t'}dt_2 \ \delta(t_1-t_2) \ \theta(t-t') \ \theta(\beta+t'-t) \nonumber\\
   &=-\frac{\lambda}{6H}\left(\frac{H^3}{4\pi^2}\right)^2\int_{t_i}^Tdt\int_{t_i}^{T'}dt' \ (t'-t_i) \ \theta(t-t') \ \theta(\beta+t'-t)\nonumber\\
   &=-\frac{\lambda}{6H}\left(\frac{H^3}{4\pi^2}\right)^2\left[\frac{1}{6} \ \theta(T-T') \ \theta(\beta+T'-T)\{\beta^3 + 3t_i(T'^2+\beta^2) - 3T(2t_i-T'+\beta)(T'+\beta) + 3t_i^2\beta \right.\nonumber\\
   &\left. \hspace{2.5cm}+  3T^2(t_i+\beta) - T^3 - 2T'^3\} + \frac{\beta}{2}(T'-t_i)^2 \ \theta(T-T') \ \theta(T - \beta - T')\right. \nonumber\\
   &\hspace{2.5cm}\left.+ \frac{\beta}{6} \ \theta(T'-T)\{\beta^2+3t_i\beta+-3T(2t_i+\beta) + 3t_i^2 + 3T^2\}\right].
   \end{align}
   Let us collect the results from (\ref{res1}) and (\ref{res2}) to find the two-point correlation of the long-wavelength fields at first order in $\lambda$. Defining $\tau = T-t_i$ and $\tau'=T'-t_i$, we obtain

   \begin{align}
       \langle\varphi(\tau)\varphi(\tau')\rangle =& -\frac{\lambda H^5}{576\pi^4}\left[\tau'^3 + 3\tau^2\tau' - 6\tau'^2\beta-6\tau'\beta^2+2\beta^3\right] \ \theta(\tau-\tau')\nonumber\\
       &-\frac{\lambda H^5}{576\pi^4}\left[\tau^3 + 3\tau\tau'^2 - 6\tau^2\beta-6\tau\beta^2+2\beta^3\right] \ \theta(\tau'-\tau)\nonumber\\
       &+\frac{\lambda H^5}{576\pi^4}\left[\tau^3-3\tau\tau'^2+2\tau'^3-3\tau^2\beta+3\tau\beta^2-\beta^3\right] \ \theta(\tau-\tau')\theta(\tau'+\beta-\tau)\nonumber\\
       &+\frac{\lambda H^5}{576\pi^4}\left[\tau'^3-3\tau^2\tau^2+2\tau^3-3\tau'^2\beta+3\tau'\beta^2-\beta^3\right] \ \theta(\tau'-\tau)\theta(\tau+\beta-\tau')\nonumber\\
       &-\frac{\lambda H^5\beta}{192\pi^4}\tau'^2 \ \theta(\tau-\tau')\theta(\tau-\tau'-\beta)\nonumber\\
       &-\frac{\lambda H^5\beta}{192\pi^4}\tau^2 \ \theta(\tau'-\tau)\theta(\tau'-\tau-\beta).
   \end{align}
   In the limit $\beta=0$ (which means $N=1$) the expression above reduces to

   \begin{align}
        \langle\varphi(\tau)\varphi(\tau')\rangle =& -\frac{\lambda H^5}{576\pi^4}(\tau'^3 + 3\tau^2\tau') \ \theta(\tau-\tau')-\frac{\lambda H^5}{576\pi^4}(\tau^3 + 3\tau\tau'^2) \ \theta(\tau'-\tau).
   \end{align}
For $\tau>\tau'$ this further reduces to

\begin{align}
    \langle\varphi(\tau)\varphi(\tau')\rangle& = -\frac{\lambda H^5}{576\pi^4}(\tau'^3+3\tau^2\tau'), 
\end{align}
which matches with the results of \cite{Onemli:2015pma} where it was shown that the stochastic formulation and the QFT calculation give the same result (See Eq.(77) for the result from QFT calculation in \cite{Onemli:2015pma}). However, here we obtain additional terms that depend on $\beta$. Most importantly these are independent of the coarse-graining parameter $\sigma$ which means they survive even when we take the limit $\sigma<<1$ at the end of the calculation. Although we know that these terms originate due to non-orthogonality of the fields $\varphi_i^N$ and $\psi_i$ as discussed in subsection \ref{4A}, their significance from a physical viewpoint is not very clear to us. Atleast from a mathematical viewpoint, we can see no reason to drop these additional terms and so we regard these as genuine contributions to the two-point correlation of the long-wavelength fields. Nevertheless this calculation clearly demonstrates that the stochastic formalism yields an expression for the two-point correlation of the long-wavelength fields that is different when compared with the QFT calculation. Also the deviation occurs at $\mathcal{O}(\lambda)$ which means the two methods give same results for the case of free theory but differ for the case of an interacting theory. 

\section{Treatment of general interaction terms in stochastic theory}\label{Sec5}

In this section we will highlight another problem that plagues the lagrangian formulation of the stochastic inflation. To do this, we shall consider other diagrams in the expression for $S_{\text{inf}}$ that necessarily arises due to interactions. The problem that we will highlight here is how to deal with the imaginary part of these diagrams. Specifically, we want to see if we can apply a transformation similar to (\ref{hs1}) and have an interpretation of the imaginary part as a random force in the equation of motion just like we did in the previous sections. To do this, we propose a procedure where we expand the imaginary part of the influence functional that contains terms that are of higher-order in the small coupling constant. In what follows we will not explicitly compute the diagrams but assume a general form of the expressions. The most general expression for the imaginary part of any diagram in $S_{\text{inf}}$ that contributes to the path integral is given by

\begin{align}
    \exp\left(-\alpha \ \varphi^{N_1}_{i_1}\varphi^{N_2}_{i_2}...\varphi^{N_M}_{i_M}B_{i_1i_2...i_M}\right)\label{genterm}.
\end{align}

For example the contribution to the path integral of the imaginary part of the second diagram in $S_{\text{inf}}$ is given by

\begin{align}
    \exp\left(-\frac{\lambda}{4}\varphi_i^2\varphi_j\varphi_k\{\eta_i\eta_j\eta_k\overrightarrow{\Lambda}_j\overrightarrow{\Lambda}_k\text{Im}(F_{ij}F_{ik})\}\right)\label{asym1}.
\end{align}

We cannot see how the transformation (\ref{hs1}) can be applied here. There is however some hope in dealing with a certain class of expressions by the use of perturbation theory. Let us assume without loss of generality that the interacting theory leads to the following contribution to the path integral in addition to terms arising from the free part:

\begin{align}
    \exp\left(-\frac{1}{2}(\lambda\varphi_i^N)B_{ij}(\lambda\varphi^N_j)\right),\hspace{2cm}\text{with}\hspace{0.3cm}N=2.\label{interact1}
\end{align}

The reason for choosing this form is that one can directly apply the Hubbard-Stratonovich transformation here so that the correlations of the long-wavelength fields can be cross checked with those from perturbation theory. Including the contribution from the free theory as well and directly applying the Hubbard-Stratonovich transformation leads to

\begin{align}
    \exp\left(-\frac{1}{2}\varphi_iA_{ij}\varphi_j\right)\exp\left(-\frac{1}{2}(\lambda\varphi_i^N)B_{ij}(\lambda\varphi^N_j)\right) &= \int\mathscr{D}\xi\int\mathscr{D}\zeta\exp\left(-\frac{1}{2}\xi_i(A^{-1}_{ij})\xi_j\right)\exp\left(-\frac{1}{2}\zeta_i(B^{-1}_{ij})\zeta_j\right)\nonumber\\
    &\hspace{4cm}\times\exp(i\xi_i\varphi_i)\exp\{i\zeta_i(\lambda\varphi_i^N)\}\label{hs3}.
\end{align}

where $\mathscr{D}\xi = \prod_i\mathscr{D}\xi_i$ and $\exp\left(-\frac{1}{2}\varphi_iA_{ij}\varphi_j\right)$ comes from the free theory and may correspond to (\ref{ft01i}). The effective action and the equations of motion (assuming slow-roll) for $N=2$ reads

\begin{align}
&S_{\text{eff}} = S_+[\varphi_+]-S_-[\varphi_-] + \xi_i\varphi_i + \lambda\zeta_i\varphi^2_i,\\
&\dot\varphi_i = \xi_i + 2\lambda\varphi_i\zeta_i.\label{eqmotion2}
\end{align}

(Note that we have for simplicity ignored any interaction part from the classical action of long-wavelength fields since we are interested here in the random part of the equation of motion.) The two-point correlation for $\varphi$ can then be obtained at $\mathcal{O}(\lambda^2)$ as

\begin{align}
    \langle\varphi_I(T)\varphi_J(T')\rangle &= 2^2\lambda^2\int^T dt\int^{T'}dt'\int^t dt_1\int^{t'}dt_2 \  \langle\xi_I(t_1)\xi_J(t_2)\rangle\times\langle\zeta_I(t)\zeta_J(t')\rangle\nonumber\\
    &=2^2\lambda^2\int^T dt\int^{T'}dt'\int^t dt_1\int^{t'} dt_2 \ A_{IJ}(t_1,t_2)B_{IJ}(t,t') \label{reshs1}.
\end{align}

Let us now propose the following perturbative approach to reproduce the result obtained above. We multiply (\ref{interact1}) by an auxiliary term followed by expanding (\ref{interact1}) in powers of $\lambda^2$:

\begin{align}
     &\exp\left(-\frac{1}{2}(\lambda\varphi_i^N)B_{ij}(\lambda\varphi^N_j)\right)\times \exp\left(-\frac{1}{2}(\varphi_i^N)C_{ij}(\varphi^N_j) + J_i\varphi_i^N\right)\nonumber\\ &\approx \left(1-\frac{1}{2}(\lambda\varphi_i^N)B_{ij}(\lambda\varphi^N_j) + \mathcal{O}(\lambda^4)\right)\exp\left(-\frac{1}{2}(\varphi_i^N)C_{ij}(\varphi^N_j)+J_i\varphi^N_i\right)\nonumber\\
     &=\left(1-\frac{\lambda^2B_{ij}}{2}\left(\frac{\delta}{\delta J_i}\right)\left(\frac{\delta}{\delta J_j}\right) + \mathcal{O}(\lambda^4)\right)\exp\left(-\frac{1}{2}(\varphi_i^N)C_{ij}(\varphi^N_j)+J_i\varphi^N_i\right)\label{aux1}.
\end{align}
In the end we must take both $C_{ij}$ and $J_i$ to be zero. Including the expression from free theory, taking $N=2$ and applying the Hubbard-Stratonovich transformation we obtain

\begin{align}
    &\left(1-\frac{\lambda^2B_{ij}}{2}\left(\frac{\delta}{\delta J_i}\right)\left(\frac{\delta}{\delta J_j}\right) + \mathcal{O}(\lambda^4)\right)\int\mathscr{D}\xi \int\mathscr{D}\zeta \ \exp\left(-\frac{1}{2}\xi_i(A^{-1}_{ij})\xi_j\right)\exp(i\xi_i\varphi_i)\nonumber\\
    &\hspace{1cm}\times \exp\left(-\frac{1}{2}\zeta_i(C^{-1}_{ij})\zeta_j+i\zeta_i\varphi_i^2-i\zeta_i(C^{-1}_{ij})J_j+\frac{1}{2}J_i(C^{-1}_{ij})J_j\right)\nonumber\\
    &=\int\mathscr{D}\zeta\left[1-\frac{\lambda^2B_{ij}}{2}\Big\{(-i\zeta_kC^{-1}_{ki})(-i\zeta_lC^{-1}_{lj})+C^{-1}_{ij}\Big\}\right]\exp\left(-\frac{1}{2}\zeta_i(C^{-1}_{ij})\zeta_j+i\zeta_i\varphi_i^2\right)\nonumber\\
    &\hspace{1cm}\times\int\mathscr{D}\xi \ \exp\left(-\frac{1}{2}\xi_i(A^{-1}_{ij})\xi_j\right)\exp(i\xi_i\varphi_i),\label{pb1}
\end{align}
where we took $J_i=0$ in the second step. We now have a noise $\zeta$ with a modified statistics given by the probability distribution

\begin{align}
   P[\zeta]= \left(1-\frac{\lambda^2B_{ij}}{2}\Big\{(-i\zeta_kC^{-1}_{ki})(-i\zeta_lC^{-1}_{lj})+C^{-1}_{ij}\Big\}\right)\exp\left(-\frac{1}{2}\zeta_i(C^{-1}_{ij})\zeta_j\right)\label{pb1a}.
\end{align}

The equation of motion is 

\begin{align}\label{eqmotion3}
    \dot\varphi_i = \xi_i + 2\varphi_i\zeta_i.
\end{align}

which is slightly different as compared with the earlier one in (\ref{eqmotion2}). The two-point correlation of the long-wavelength fields can be calculated as

\begin{align}
    \langle\varphi_I(T)\varphi_J(T')\rangle = \int^Tdt\int^{T'}dt' \ (\langle\xi_I(t)\xi_J(t')\rangle + 2^2\langle\xi_I(t)\xi_J(t')\rangle\langle\zeta_I(t)\zeta_J(t')\rangle).
\end{align}

The $\mathcal{O}(\lambda^2)$ contribution comes from the $\langle\zeta(t)\zeta(t')\rangle$ which reads

\begin{align}
    \langle\zeta_i\zeta_j\rangle &= C_{ij}-\frac{\lambda^2}{2}B_{kl}C_{ij}C^{-1}_{kl}+\frac{\lambda^2}{2}B_{kl}C^{-1}_{lm}C^{-1}_{kn}(C_{mi}C_{nj}+C_{mj}C_{ni}+C_{mn}C_{ij})\nonumber\\
    &=C_{ij}+\lambda^2B_{ij}.
\end{align}
Taking $C_{ij}=0$ we obtain

\begin{align}
    \langle\zeta_i\zeta_j\rangle = \lambda^2B_{ij}.
\end{align}
Thus, the two-point correlation of the long-wavelength fields at $\mathcal{O}(\lambda^2)$ is

\begin{align}
    \langle\varphi_I(T)\varphi_J(T')\rangle = 2^2\lambda^2\int^Tdt\int^{T'}dt'\int^tdt_1\int^{t'}dt_2 \ A_{IJ}(t_1,t_2)B_{IJ}(t_1,t_2)\label{respert1}
\end{align}

which matches with (\ref{reshs1}). But, we cannot go beyond two-point function using the new probability distribtuion (\ref{pb1})because one must expand (\ref{aux1}) beyond $\mathcal{O}(\lambda^2)$. Thus, the two methods agree as long as we expand (\ref{aux1}) in power of $\lambda^2$ up to an appropriate order. But what would happen if we use a different auxiliary function? Let us begin with the following expression in place of (\ref{aux1}):

\begin{align}
     &\exp\left(-\frac{1}{2}(\lambda\varphi_i^N)B_{ij}(\lambda\varphi^N_j)\right)\times \exp\left(-\frac{1}{2}(\varphi_i)C_{ij}(\varphi_j) + J_i\varphi_i\right)\nonumber\\ &\approx \left(1-\frac{1}{2}(\lambda\varphi_i^N)B_{ij}(\lambda\varphi^N_j) + \mathcal{O}(\lambda^4)\right)\exp\left(-\frac{1}{2}(\varphi_i)C_{ij}(\varphi_j)+J_i\varphi_i\right)\nonumber\\
     &=\left(1-\frac{\lambda^2B_{ij}}{2}\left(\frac{\delta}{\delta J_i}\right)^N\left(\frac{\delta}{\delta J_j}\right)^N + \mathcal{O}(\lambda^4)\right)\exp\left(-\frac{1}{2}(\varphi_i)C_{ij}(\varphi_j)+J_i\varphi_i\right).\label{aux2}
\end{align}
Note that the chosen auxiliary function is of the same form as that of the free part and therefore this procedure is similar to what we do in usual perturbation theory when dealing with the interaction terms of the real action in QFT. Including the contribution from free part, taking $N=2$ and applying the Hubbard-Stratonovich transformation we obtain

\begin{align}
    &\left(1-\frac{\lambda^2B_{ij}}{2}\left(\frac{\delta}{\delta J_i}\right)^2\left(\frac{\delta}{\delta J_j}\right) ^2+ \mathcal{O}(\lambda^4)\right)\int\mathscr{D}\xi \int\mathscr{D}\zeta \ \exp\left(-\frac{1}{2}\xi_i(A^{-1}_{ij})\xi_j\right)\exp(i\xi_i\varphi_i)\nonumber\\
    &\hspace{1cm}\times \exp\left(-\frac{1}{2}\zeta_i(C^{-1}_{ij})\zeta_j+i\zeta_i\varphi_i-i\zeta_i(C^{-1}_{ij})J_j+\frac{1}{2}J_i(C^{-1}_{ij})J_j\right)\nonumber\\
    &=\int\mathscr{D}\zeta\left[1-\frac{\lambda^2B_{ij}}{2}\delta_{ik}\delta_{jl}\Big\{(-i\zeta_mC^{-1}_{mi})(-i\zeta_nC^{-1}_{nj})(-i\zeta_pC^{-1}_{pk})(-i\zeta_qC^{-1}_{ql})+(-i\zeta_nC^{-1}_{nk})(-i\zeta_nC^{-1}_{nl})C^{-1}_{ij}\right.\nonumber\\
    &\left.\hspace{2cm}+(-i\zeta_nC^{-1}_{nj})(-i\zeta_nC^{-1}_{nl})C^{-1}_{ik}+(-i\zeta_nC^{-1}_{nj})(-i\zeta_nC^{-1}_{nk})C^{-1}_{il}+(-i\zeta_nC^{-1}_{ni})(-i\zeta_nC^{-1}_{nl})C^{-1}_{jk}\right.\nonumber\\
    &\left.\hspace{2cm}+(-i\zeta_nC^{-1}_{ni})(-i\zeta_nC^{-1}_{nk})C^{-1}_{jl}+(-i\zeta_nC^{-1}_{ni})(-i\zeta_nC^{-1}_{nj})C^{-1}_{kl}+C^{-1}_{ij}C^{-1}_{kl}+C^{-1}_{ik}C^{-1}_{jl}+C^{-1}_{il}C^{-1}_{jk}\Big\}\right]\nonumber\\
    &\hspace{2cm}\times\exp\left(-\frac{1}{2}\zeta_i(C^{-1}_{ij})\zeta_j+i\zeta_i\varphi_i\right)\times\int\mathscr{D}\xi \ \exp\left(-\frac{1}{2}\xi_i(A^{-1}_{ij})\xi_j\right)\exp(i\xi_i\varphi_i),\label{pb2}
\end{align}

where we took $J_i=0$ in the second step. Now we have a noise $\zeta_i$ with a different probability distribution. The equation of motion is also different:

\begin{align}\label{eqmotion4}
    \dot\varphi_i = \xi_i+\zeta_i.
\end{align}

 If we calculate two-point correlation of $\varphi$ we obtain

 \begin{align}
     \langle\varphi_I(T)\varphi_J(T')\rangle = \int^Tdt\int^{T'}dt'(\langle\xi_I(t)\xi_J(t')\rangle + \langle\zeta_I(t)\zeta_J(t')\rangle.
 \end{align}
 
 However, this gives a different expression because $\langle\zeta_i\zeta_j\rangle = 0$ for the probability distribution given in (\ref{pb2}). The $\mathcal{O}(\lambda^2)$ contribution which is missing in the two-point correlation here can be obtained only if we consider the four-point correlation of $\varphi$. It is clear that although both the methods seems mathematically consistent, they give different results for the correlations of the long-wavelength fields. On the bright side, choosing the auxiliary function as in (\ref{aux2}) naturally does not give rise to additional $\mathcal{O}(\lambda)$ terms that are dependent on $\beta$ in the two-point correlation of the long-wavelength fields and hence it gives the same results as obtained in QFT calculations. However, we cannot say for sure that the auxiliary function used in (\ref{aux2}) is the correct one unless we verify the results for higher-point correlations as well. Even if it can be verified that the auxiliary function chosen in (\ref{aux2}) is correct, an ensuing question would arise as to why one must not directly apply the HS transformation as done in (\ref{hs2}).
 
 Therefore, the analysis that we have carried out prompts us to follow either of the two procedures: 
 
 \begin{enumerate}
     \item For symmetric terms like (\ref{interact1}) one must directly apply the HS transformation as done in (\ref{hs3}) but for assymetric terms like those given (\ref{asym1}) and (\ref{genterm}) one must follow perturbation theory using the free part itself as the auxiliary function. This is because the procedure would become more in line with the usual perturbation theory we use to deal with the interaction terms of the real action in QFT. In this light, we can say that when it comes to dealing with imaginary action, terms like (\ref{interact1}) behave like kinetic terms since one can directly apply HS transformation just like we do for the free part (\ref{ft01i}). The ones which truly represent interaction terms are like those given (\ref{asym1}) where one cannot directly apply the HS transformation. The problem with this procedure is how to interpret the additional terms arising at $\mathcal{O}(\lambda)$ from a physical viewpoint since they don't arise in QFT calculations. 

     \item Apply perturbation theory to any general expression of the form given in (\ref{genterm}) using the free part itself as the auxiliary function. This procedure yields a result for the two-point correlation that matches with that from QFT calculations although it remains to be seen whether the results match for higher-point correlations as well. The problem with this procedure is that it raises the question of why the HS transformation that is allowed from a mathematical viewpoint should not be applied directly to the symmetric terms as done in (\ref{hs3}).
     
 \end{enumerate}
 It will be crucial to address these issues to confirm the viability of the lagrangian formulation of stochastic inflation.

\section{Conclusion}\label{Conclusion}

 In this paper, we reviewed the lagrangian formulation of stochastic inflation and certain dilemmas associated with its application in deriving correlation functions in an interacting theory. We used this formalism to derive the influence functional that contains all the effects due to the short-wavelength fields. These effects influence the dynamics of the long-wavelength fields through the influence functional that appears in the effective action of the long-wavelength fields. Of particular interest was the imaginary contribution of this influence functional to the effective action. For free theory, it becomes possible to interpret the imaginary part as a stochastic force in the equation of motion of the long-wavelength fields. This is achieved through the application of Hubbard-Stratonovich transformation that re-introduces a path integral in the reduced generating functional over classical random fields that follow a certain probability distribution function. In this way, the quantum expectation values of the long-wavelength fields turned into statistical averages. This method of computing correlation functions of the long-wavelength fields yielded results that are identical to those calculated using the QFT technique, at least for the case of a free theory. In the past literature \cite{Onemli:2015pma, Tsamis:2005hd}, this method was shown to work for interacting theory as well, however, as shown in section \ref{4D}, we obtain additional terms in the two-point correlation of the long-wavelength fields that does not appear in QFT \cite{Tsamis:2005hd, Onemli:2015pma} nor appears in a similar work that uses the lagrangian formluation of stochastic inflation \cite{PerreaultLevasseur:2013kfq}. An important feature of the additional terms is that they are independent of the coarse-graining parameter. The reason these additional terms appear was due to the fact that $\varphi_i^N$ and $\psi_i$ are not strictly orthogonal for $N>1$. Although the physical significance of these additional terms is lacking we found no reason to avoid these from a mathematical viewpoint.
 
 Further dilemma is caused when we considered general expressions of the form (\ref{genterm}) that could arise in an interacting theory. We observed that for a special class of terms that are symmetric (\ref{interact1}), one can either directly apply the HS transformation or use perturbation theory as proposed in section \ref{Sec5} to derive the effective action. The results for the two-point correlation of the long-wavelength fields derived by using these effective actions matched only if we used a particular form of the auxiliary function. Furthermore, we observed that if we used a different auxiliary function such as the one given in (\ref{aux2}) that has a form similar to that of the free part, then the results differ. The analysis left open two ways to follow, both of which have been found to be unreliable. 1) One can either apply the HS transformation directly on symmetric terms such as (\ref{interact1}) and deal with more general terms such as (\ref{genterm}) by using perturbation theory. Although it is not really clear which auxiliary function to use, the only sensible choice is that of the free part itself, because it would be more in line with the usual perturbation theory. The caveat of this procedure is that additional terms arise, such as those calculated in (\ref{4D}). These do not appear in standard QFT results, and therefore their significance is questionable. 2) One can simply apply perturbation to any term of the form (\ref{genterm}) by using the free part itself as the auxiliary function. The advantage of this method is two-fold. One, that this way of doing perturbation theory is more in line with the standard way, and second, no controversial terms arise at least in the two-point correlation of the long-wavelength fields. The caveat here is that it makes it unclear why are we not allowed to apply the HS transformation directly to symmetric terms such as those given in (\ref{interact1}). We saw no reason why this is not allowed, at least from a mathematical point of view.

 The analysis carried out in this paper therefore points towards a general problem associated with the interpretation of the imaginary parts of the effective action. It turns out that the treatment of the imaginary parts needs a different approach from that of the real parts as we have made clear in either of the procedures that one could follow. A deeper look into the issues we have raised in this paper is necessary to fully confirm the viability of the lagrangian formulation of stochastic inflation. 

\acknowledgments
We thank Laurence Perreault Levasseur for valuable discussions and Ebin Paul for preliminary calculations that helped motivate this work. The work of R. K. P. was supported by UGC(India).

\appendix
\section{Green's functions}\label{Appendix A}
Green's functions for free theory in CTP formalism is given as
\begin{align}
    & G_{++}^{(0)}(x,y)=-i \langle \hat{T} (\psi(x)\psi(y)\rangle =-i\int \frac{d^3x}{(2\pi)^3}W(k,t_x)W(k,t_y) \exp{[-i \Vec{k}.(\Vec{x}-\Vec{y})]} \{ \theta(t_x-t_y) \phi_k(t_x) \phi_k^*(t_y) \nonumber\\
    &\hspace{5cm} + \theta(t_y-t_x) \phi_{-k}(t_y) \phi_{-k}^*(t_x)\} \\
    & G_{+-}^{(0)}(x,y)=-i \langle(\psi(y)\psi(x)\rangle\hspace{0.23cm}=-i\int \frac{d^3x}{(2\pi)^3}W(k,t_x)W(k,t_y) \exp{[-i \Vec{k}.(\Vec{x}-\Vec{y})] \phi_{-k}(t_y)\phi_{-k}^*(t_x) }\\
    & G_{-+}^{(0)}(x,y)=-i \langle(\psi(x)\psi(y)\rangle\hspace{0.23cm}=-i\int \frac{d^3x}{(2\pi)^3}W(k,t_x)W(k,t_y) \exp{[-i \Vec{k}.(\Vec{x}-\Vec{y}) \phi_k(t_x) \phi_k^*(t_y) ]}\\
    & G_{--}^{(0)}(x,y)=-i \langle \hat{\Bar{T}} (\psi(x)\psi(y)\rangle=-i\int \frac{d^3x}{(2\pi)^3}W(k,t_x)W(k,t_y) \exp[-i \Vec{k}.(\Vec{x}-\Vec{y})]\{ \theta(t_y-t_x) \phi_k(t_x) \phi_k^*(t_y)\nonumber\\
    &\hspace{5cm} + \theta(t_x-t_y) \phi_{-k}(t_y) \phi_{-k}^*(t_x)\}.
\end{align}

We may also require a shorter way of writing the Green's functions as follows:

\begin{align}
      & G_{++}^{(0)}(x,y)=-i \langle \hat{T} (\psi(x)\psi(y)\rangle =-i\int \frac{d^3x}{(2\pi)^3}W(k,t_x)W(k,t_y) \exp{[-i \Vec{k}.(\Vec{x}-\Vec{y})]} f_{++}(\mathbf{k},t_x,t_y)\\
    & G_{+-}^{(0)}(x,y)=-i \langle(\psi(y)\psi(x)\rangle\hspace{0.23cm}=-i\int \frac{d^3x}{(2\pi)^3}W(k,t_x)W(k,t_y) \exp[-i \Vec{k}.(\Vec{x}-\Vec{y})] f_{+-}(\mathbf{k},t_x,t_y)\\
    & G_{-+}^{(0)}(x,y)=-i \langle(\psi(x)\psi(y)\rangle\hspace{0.23cm}=-i\int \frac{d^3x}{(2\pi)^3}W(k,t_x)W(k,t_y) \exp[-i \Vec{k}.(\Vec{x}-\Vec{y})] f_{-+}(\mathbf{k},t_x,t_y) \\
    & G_{--}^{(0)}(x,y)=-i \langle \hat{\Bar{T}} (\psi(x)\psi(y)\rangle=-i\int \frac{d^3x}{(2\pi)^3}W(k,t_x)W(k,t_y) \exp[-i \Vec{k}.(\Vec{x}-\Vec{y})]f_{--}(\mathbf{k},t_x,t_y).  
\end{align}

where

\begin{align}
    &f_{++}(\mathbf{k},t_x,t_y) = \theta(t_x-t_y) \phi_k(t_x) \phi_k^*(t_y) + \theta(t_y-t_x) \phi_{-k}(t_y) \phi_{-k}^*(t_x)\\
    &f_{+-}(\mathbf{k},t_x,t_y) = \phi_{-k}(t_y) \phi_{-k}^*(t_x)\\
    &f_{-+}(\mathbf{k},t_x,t_y) = \phi_k(t_x) \phi_k^*(t_y)\\
    &f_{--}(\mathbf{k},t_x,t_y) = \theta(t_x-t_y)\phi_{-k}(t_y) \phi_{-k}^*(t_x)  + \theta(t_y-t_x) \phi_k(t_x) \phi_k^*(t_y)
\end{align}

We define $G_{ij}^{(0)} = -iF_{ij}$. Our goal is to determine the properties of the real and imaginary components of $F_{ij}$. Focusing first on the real part, one can write for $F_{++}$ ,\\
    \begin{align}
      &  \text{Re} F_{++} =  \text{Re} \int \frac{d^3x}{(2\pi)^3}W(k,t_x)W(k,t_y) \exp{[-i \Vec{k}.(\Vec{x}-\Vec{y})]} \{ \theta(t_x-t_y) \phi_k(t_x) \phi_k^*(t_y) + \theta(t_y-t_x) \phi_{-k}(t_y) \phi_{-k}^*(t_x)\} \nonumber\\
      &\hspace{1.1cm}= \int \frac{d^3x}{(2\pi)^3}W(k,t_x)W(k,t_y) \frac{1}{2}\left\{ \exp{[-i \Vec{k}.(\Vec{x}-\Vec{y})]} \{ \theta(t_x-t_y) \phi_k(t_x) \phi_k^*(t_y) + \theta(t_y-t_x) \phi_{-k}(t_y) \phi_{-k}^*(t_x)\right\}\nonumber\\
      & \hspace{3cm}
     \left\{ \exp{[-i \Vec{k}.(\Vec{x}-\Vec{y})]} \{ \theta(t_x-t_y) \phi_k^*(t_x) \phi_k(t_y) + \theta(t_y-t_x) \phi_{-k}^*(t_y) \phi_{-k}(t_x)\right\}.\\\nonumber
     \end{align}
Changing $\Vec{k}$ to $-\Vec{k}$ in the second term, we find
\begin{align}
      &  \text{Re} F_{++} =\int \frac{d^3x}{(2\pi)^3}W(k,t_x)W(k,t_y) \exp{[-i \Vec{k}.(\Vec{x}-\Vec{y})]} \frac{(\phi_k(t_x) \phi_k^*(t_y)+\phi_{-k}^*(t_x) \phi_{-k}(t_y))}{2}.
\end{align}
Here, by defining
  \begin{align}
      f_{++}^{Re}=\frac{1}{2}\left\{\phi_{k}(t_x) \phi_{k}^*(t_y)+\phi_{-k}^*(t_x) \phi_{-k}(t_y)\right\},
  \end{align}
one can get 
  \begin{align}
      \text{Re} F_{++} =\int \frac{d^3x}{(2\pi)^3}\exp{[-i \Vec{k}.(\Vec{x}-\Vec{y})]} W(k,t_x)W(k,t_y) f_{++}^{Re}.
  \end{align}
 The same analysis can be done for real parts of all four $F_{ij}$ terms to get the relation,
 \begin{align}
   & \text{Re} F_{++}=\text{Re}F_{+-}=\text{Re}F_{-+}=\text{Re}F_{--} ,
 \end{align}
 with all the four $F_{ij}$ terms having the same value of $f^{Re}= f_{++}^{Re}$.\\
 Now, coming to imaginary parts, for $F_{++}$,it's
 \begin{align} \label{A1f}
     &\text{Im} F_{++} =  \text{Im} \int\frac{d^3x}{(2\pi)^3}W(k,t_x)W(k,t_y) \exp{[-i \Vec{k}.(\Vec{x}-\Vec{y})]} \{ \theta(t_x-t_y) \phi_k(t_x) \phi_k^*(t_y) \nonumber\\ 
     &\hspace{2cm}+ \theta(t_y-t_x) \phi_{-k}(t_y) \phi_{-k}^*(t_x)\}\nonumber\\
    & \hspace{1.15cm}= \int \frac{d^3x}{(2\pi)^3}W(k,t_x)W(k,t_y) \frac{1}{2}\left[\theta(t_x-t_y)\left\{\exp{[-i \Vec{k}.(\Vec{x}-\Vec{y})]}\phi_k(t_x) \phi_k^*(t_y) - \exp{[i \Vec{k}.(\Vec{x}-\Vec{y})]}  \phi_{k}^*(t_x) \phi_{k}(t_y)\right\}\right. \nonumber\\
    &\hspace{2cm}\left.+
     \theta(t_y-t_x)\left\{\exp{[-i \Vec{k}.(\Vec{x}-\Vec{y})]}\phi_{-k}(t_y) \phi_{-k}^*(t_x) - \exp{[+i \Vec{k}.(\Vec{x}-\Vec{y})]}  \phi_{-k}^*(t_y) \phi_{-k}(t_x)\right\}\right] \nonumber \\
    &\hspace{1.15cm} = \int \frac{d^3x}{(2\pi)^3}W(k,t_x)W(k,t_y) \frac{1}{2} \left[\exp{[-i \Vec{k}.(\Vec{x}-\Vec{y})]} \phi_{-k}^*(t_x) \phi_{-k}(t_y)\left\{ \theta(t_y-t_x)-\theta(t_x-t_y\right\}\right. \nonumber\\
    &\hspace{2cm}\left.+ 
     \exp{[-i \Vec{k}.(\Vec{x}-\Vec{y})]} \phi_{k}(t_x) \phi_
     {k}^*(t_y)\left\{\theta(t_x-t_y)-\theta(t_y-t_x)\right\}
     \right] \nonumber\\
    & \implies \text{Im} F_{++} = \frac{1}{2} \int \frac{d^3x}{(2\pi)^3}W(k,t_x)W(k,t_y) \exp{[-i \Vec{k}.(\Vec{x}-\Vec{y})]} \left\{\theta(t_x-t_y)-\theta(t_y-t_x)\right\} \nonumber \\
    & \hspace{3.5cm} [\phi_{k}(t_x) \phi_{k}^*(t_y)- \phi_{-k}^*(t_x) \phi_{-k}(t_y)].
    \end{align}
     We also define
  
  \begin{align}
      f_1=\frac{1}{2}\left\{\theta(t_x-t_y)-\theta(t_y-t_x)\right\}[\phi_{k}(t_x) \phi_{k}^*(t_y)- \phi_{-k}^*(t_x) \phi_{-k}(t_y)],
  \end{align}
  due to which Eq.(\ref{A1f}) takes the form
  \begin{align}
        \text{Im} F_{++} = \int \frac{d^3x}{(2\pi)^3}W(k,t_x)W(k,t_y) \exp{[-i \Vec{k}.(\Vec{x}-\Vec{y})]} f_1.
  \end{align}
 
    Now coming to imaginary part of the term  $F_{--}$,
    \begin{align}
         & \text{Im} F_{--} = \text{Im} \int \frac{d^3x}{(2\pi)^3}W(k,t_x)W(k,t_y) \exp{[-i \Vec{k}.(\Vec{x}-\Vec{y})]\{ \theta(t_y-t_x) \phi_k(t_x) \phi_k^*(t_y) + \theta(t_x-t_y) \phi_{-k}(t_y) \phi_{-k}^*(t_x)\}}.
    \end{align}
    Proceeding exactly the same way, one can find that 
    \begin{align}
       & \text{Im} F_{--} = - \frac{1}{2} \int \frac{d^3x}{(2\pi)^3}W(k,t_x)W(k,t_y) \exp{[-i \Vec{k}.(\Vec{x}-\Vec{y})]} \left\{\theta(t_x-t_y)-\theta(t_y-t_x)\right\} \nonumber \\
       & \hspace{3cm}[\phi_{k}(t_x) \phi_{k}^*(t_y)- \phi_{-k}^*(t_x) \phi_{-k}(t_y)]\\
        & \implies \text{Im} F_{--}  =
       -\int \frac{d^3x}{(2\pi)^3}W(k,t_x)W(k,t_y) \exp{[-i \Vec{k}.(\Vec{x}-\Vec{y})]} f_1.
    \end{align}
    
    For imaginary part of the term $F_{+-}$, it's
     \begin{align}
          & \text{Im} F_{+-} =\text{Im} \int \frac{d^3x}{(2\pi)^3}W(k,t_x)W(k,t_y) \exp{[-i \Vec{k}\cdot(\Vec{x}-\Vec{y})]}  \phi_{-k}(t_y) \phi_{-k}^*(t_x) \nonumber\\
          & \hspace{1.1cm} = \frac{1}{2}  \int \frac{d^3x}{(2\pi)^3}W(k,t_x)W(k,t_y) \exp{[-i \Vec{k}\cdot(\Vec{x}-\Vec{y})]}[\phi_{-k}^*(t_x) \phi_{-k}(t_y)-\phi_{k}(t_x) \phi_{k}^*(t_y)] .
     \end{align}
     Again, we define 
     \begin{align}
      f_2=\frac{1}{2}[ \phi_{-k}^*(t_x) \phi_{-k}(t_y) - \phi_{k}(t_x) \phi_{k}^*(t_y)]
    \end{align}
     \begin{align}
        \implies \text{Im} F_{+-} = \int \frac{d^3x}{(2\pi)^3}W(k,t_x)W(k,t_y) \exp{[-i \Vec{k}.(\Vec{x}-\Vec{y})]} f_2.
  \end{align}

     And finally for imaginary part of $F_{-+}$, it's
     \begin{align}
          & \text{Im} F_{-+} =\text{Im} \int \frac{d^3x}{(2\pi)^3}W(k,t_x)W(k,t_y) \exp{[-i \Vec{k}\cdot(\Vec{x}-\Vec{y})]}  \phi_{k}(t_y) \phi_{k}^*(t_x) \nonumber\\
          & \hspace{1.1cm} = \frac{1}{2}  \int \frac{d^3x}{(2\pi)^3}W(k,t_x)W(k,t_y) \exp{[-i \Vec{k}\cdot(\Vec{x}-\Vec{y})]}[\phi_{k}(t_x) \phi_{k}^*(t_y)-\phi_{-k}^*(t_x) \phi_{-k}(t_y)] 
        \end{align}
     \begin{align}
        \implies \text{Im} F_{-+} = -\int \frac{d^3x}{(2\pi)^3}W(k,t_x)W(k,t_y) \exp{[-i \Vec{k}.(\Vec{x}-\Vec{y})]} f_2.
     \end{align}
     Comparing these four expressions, we find that they have the relations
      \begin{align}
    &\text{Im} F_{++}=-\text{Im}F_{--} \hspace{0.3cm},\\
    &\text{Im}F_{+-}=-\text{Im}F_{-+} \hspace{0.3cm}.\\ \nonumber
 \end{align}

\section{1-point derivatives and 2 point derivatives }\label{Appendix B}
\subsection{1-point derivatives}
\begin{align}
&\left(\frac{1}{i\eta_i} \frac{\delta}{\delta J_i}\right) \tilde{Z}_{\text{f}}[\varphi_\pm;J_\pm]=  
\right)  \tilde{Z}_{\text{f}}[\varphi_\pm;J_\pm].
    \end{align}

 \subsection{2-point derivatives}
\begin{align}
    &\left(\frac{1}{i\eta_i} \frac{\delta}{\delta J_i}\right) \left(\frac{1}{i\eta_j} \frac{\delta}{\delta J_j}\right)\tilde{Z}_{\text{f}}[\varphi_\pm;J_\pm]=  \left( %
\right)\tilde{Z}_{\text{f}}[\varphi_\pm;J_\pm]. 
\end{align}  
\section{Noise correlations}\label{Appendix C}
Let us perform the calculation of the various two-point noise correlations.

\begin{align}
    \langle\xi_1(x)\xi_1(y)\rangle &= \int \frac{d^3\mathbf{k}}{(2\pi)^3} A_{11}(\mathbf{k},t_x,t_y)\nonumber\\
    &=\int \frac{dk \ k^2}{(2\pi)^3}\int_0^\pi\sin\theta \ d\theta\int_0^{2\pi}d\phi  \ \frac{H^2}{2k^3}(-kH\delta(k-\sigma a_xH))(-kH\delta(k-\sigma a_yH)) \nonumber\\
    &\hspace{4cm} \times\left[\cos\left\{\frac{k}{H}\left(\frac{1}{a_x}-\frac{1}{a_y}\right)-\mathbf{k}\cdot(\mathbf{x}-\mathbf{y})\right\}\left(1+\frac{k^2}{H^2a_xa_y}\right) \right.\nonumber\\
    &\left.\hspace{5cm}+ \frac{k}{H}\left(\frac{1}{a_x}-\frac{1}{a_y}\right)\sin\left\{\frac{k}{H}\left(\frac{1}{a_x}-\frac{1}{a_y}\right)-\mathbf{k}\cdot(\mathbf{x}-\mathbf{y})\right\}\right]. 
\end{align}
Since we are interested in two-point correlations of long-wavelength fields at coincident comoving spatial locations we take $|\mathbf{x}-\mathbf{y}| = 0$ resulting in the following simplification of the two-point noise correlation

\begin{align}
    \langle\xi_1(x)\xi_1(y)\rangle &= \frac{H^4k}{4\pi^2}\delta(k-\sigma a_yH)\left(\cos\left\{\sigma\left(1-\frac{a_x}{a_y}\right)\right\}\left(1+\frac{\sigma^2a_x}{a_y}\right) + \sigma\left(1-\frac{a_x}{a_y}\right)\sin\left\{\sigma\left(1-\frac{a_x}{a_y}\right)\right\}\right)\Bigg|_{k=\sigma a_xH}\nonumber\\
    &\approx\frac{H^3}{4\pi^2}\delta(t_x-t_y) + \mathcal{O}(\sigma),
\end{align}
where we have kept only those terms that survive in the limit $\sigma=0.$

Next one is

\begin{align}
    \langle\xi_1(x)\xi_2(y)\rangle &= \int \frac{d^3\mathbf{k}}{(2\pi)^3} A_{12}(\mathbf{k},t_x,t_y)\nonumber\\
    &=\int \frac{dk \ k^2}{(2\pi)^3}\left(\frac{k}{\sigma a_y H}\right)\int_0^\pi\sin\theta \ d\theta\int_0^{2\pi}d\phi  \ \frac{H^2}{2k^3}(-kH\delta(k-\sigma a_xH))(-kH\delta(k-\sigma a_yH)) \nonumber\\
    &\hspace{4cm} \times\left[\cos\left\{\frac{k}{H}\left(\frac{1}{a_x}-\frac{1}{a_y}\right)-\mathbf{k}\cdot(\mathbf{x}-\mathbf{y})\right\}\left(1+\frac{k^2}{H^2a_xa_y}\right) \right.\nonumber\\
    &\left.\hspace{5cm}+ \frac{k}{H}\left(\frac{1}{a_x}-\frac{1}{a_y}\right)\sin\left\{\frac{k}{H}\left(\frac{1}{a_x}-\frac{1}{a_y}\right)-\mathbf{k}\cdot(\mathbf{x}-\mathbf{y})\right\}\right]. 
\end{align}

Repeating the same steps as before we obtain

\begin{align}
      \langle\xi_1(x)\xi_2(y)\rangle = \frac{\sigma H^3}{4\pi^2}\delta(t_x-t_y) + \mathcal{O}(\sigma^2).
\end{align}
 Thus, this correlation begins at first order in $\sigma$ and so we drop it entirely in the calculation of correlation functions of long-wavelength fields.

 A similar calculation leads to

 \begin{align}
     \langle\xi_2(x)\xi_2(y)\rangle = \frac{\sigma^2H^3}{4\pi^3}\delta(t_x-t_y) + \mathcal{O}(\sigma^3).
 \end{align}

 So we drop this correlation function as well. This implies that we must drop $\xi_2$ altogether in the equation of motion (\ref{eqmotion1}). Now let us look at other ones such as

 \begin{align}
     \langle\dot\xi_1(x)\xi_1(y)\rangle = \frac{d}{dt_x}\left(\frac{H^3}{4\pi^2}\delta(t_x-t_y)+\mathcal{O}(\sigma)\right).
 \end{align}

 This term does not contribute when we calculate the two-point correlation of long-wavelength fields. This can be seen as follows. To obtain the two-point correlation of long-wavelength fields we require time integrals of the two-point noise correlations such as that given below

 \begin{align}
     \int^T_{t_i}dt_x\int^{T'}_{t_i}dt_y\frac{d}{dt_x}\langle\dot\xi_1(x)\xi_1(y)\rangle =  \int^T_{t_i}dt_x\int^{T'}_{t_i}dt_y\frac{d}{dt_x}\left(\frac{H^3}{4\pi^2}\delta(t_x-t_y)+\mathcal{O}(\sigma)\right),
 \end{align}
 which yields zero when integrated over $t_y$. In a similar manner other correlations such as $\langle\dot\xi_1(x)\dot\xi_1(y)\rangle$ and $\langle\dot\xi_1(x)\xi_2(y)\rangle$ can also be ignored. As a result we can throw away the $\dot\xi_1$ term from the equation of motion (\ref{eqmotion1}). 

\bibliography{biblio.bib}

\end{document}